\documentclass[11pt]{amsart}
\usepackage{amsmath}
\usepackage{amsxtra}
\usepackage{amscd}
\usepackage{amsthm}
\usepackage{amsfonts}
\usepackage{amssymb}
\usepackage{eucal}

\usepackage{latexsym,amsthm}
\usepackage[english]{babel}

\DeclareFontEncoding{OT2}{}{}

\newtheorem{thm}{Theorem}[section]

\theoremstyle{remark}
\newtheorem{remark}[thm]{Remark}

\theoremstyle{definition}

\numberwithin{equation}{section}

\newcommand{\bean}{\begin{eqnarray}}
\newcommand{\eean}{\end{eqnarray}}
\newcommand{\be}{\begin{displaymath}}
\newcommand{\ee}{\end{displaymath}}
\newcommand{\bea}{\begin{eqnarray*}}   
\newcommand{\eea}{\end{eqnarray*}}

\newcommand{\secref}[1]{Section~\ref{#1}}

\newcommand{\remref}[1]{Remark~\ref{#1}}

\newcommand{\nc}{\newcommand}
\nc{\on}{\operatorname}

\nc{\ch}{\mbox{ch}}
\nc{\Z}{{\mathbb Z}}
\nc{\C}{{\mathbb C}}
\nc{\pone}{{\mathbb P}^1}
\nc{\pa}{\partial}
\nc{\F}{{\mathcal F}}
\nc{\arr}{\rightarrow}
\nc{\larr}{\longrightarrow}
\nc{\al}{\alpha}
\nc{\ri}{\rangle}
\nc{\lef}{\langle}
\nc{\W}{{\mathcal W}}
\nc{\la}{\lambda}
\nc{\ep}{\epsilon}
\nc{\su}{\widehat{{\mathfrak s}{\mathfrak l}}_2}
\nc{\sw}{{\mathfrak s}{\mathfrak l}}
\nc{\g}{{\mathfrak g}}
\nc{\h}{{\mathfrak h}}
\nc{\n}{{\mathfrak n}}
\nc{\N}{\widehat{\n}}
\nc{\G}{\widehat{\g}}
\nc{\De}{\Delta}
\nc{\gt}{\widetilde{\g}}
\nc{\Ga}{\Gamma}
\nc{\one}{{\mathbf 1}}
\nc{\z}{{\mathfrak Z}}
\nc{\La}{\Lambda}
\nc{\wt}{\widetilde}
\nc{\wh}{\widehat}
\nc{\cri}{_{\kappa_c}}
\nc{\kk}{h^\vee}
\nc{\sun}{\widehat{\sw}_N}
\nc{\si}{\sigma}
\nc{\el}{\ell}
\nc{\bi}{\bibitem}
\nc{\om}{\omega}
\nc{\ol}{\overline}
\nc{\ds}{\displaystyle}
\nc{\dzz}{\frac{dz}{z}}
\nc{\Res}{\on{Res}}
\nc{\mc}{\mathcal}
\nc{\Cal}{\mathcal}
\nc{\bb}{{\mathfrak b}}
\nc{\ot}{\otimes}
\nc{\R}{{\mathbb R}}
\nc{\yy}{{\mc Y}}
\nc{\ga}{\gamma}

\nc{\us}{\underset}
\nc{\opl}{\oplus}
\nc{\beq}{\begin{equation}}
\nc{\Fq}{{\mathcal F}}
\nc{\Mq}{{\mathcal M}}
\nc{\Rep}{\on{Rep}}
\nc{\sssec}{\subsubsection}
\nc{\ssec}{\subsection}
\nc{\lan}{\langle}
\nc{\ran}{\rangle}

\nc{\D}{\mathcal D}
\nc{\Vect}{\on{Vect}}
\nc{\ghat}{\G}
\nc{\T}{\mc T}
\nc{\Tloc}{\T^\g_{\on{loc}}}
\nc{\vac}{|0\ran}
\nc{\Wick}{{\mb :}}
\nc{\mb}{\mathbf}
\nc{\delz}{\partial_z}
\nc{\K}{{\cali K}}
\nc{\cali}{\mathcal}
\nc{\li}{\mathfrak l}
\nc{\lt}{\widetilde{\li}}
\nc{\astar}{a^*}
\nc{\cA}{{\mc A}}
\nc{\ka}{\kappa}

\nc{\OO}{{\mc O}}
\nc{\AutO}{\on{Aut} O}
\nc{\DerO}{\on{Der} O}
\nc{\DerpO}{\on{Der}_+ O}
\nc{\Au}{{\mc A}ut}
\nc{\mf}{\mathfrak}
\nc{\V}{{\mc V}}
\nc{\hh}{\wh{\h}}

\nc{\pp}{{\mathfrak p}}
\nc{\mm}{{\mathfrak m}}
\nc{\rr}{{\mathfrak r}}
\nc{\ket}{\rangle}
\nc{\zz}{{\mathfrak z}}
\nc{\gr}{\on{gr}}
\nc{\Spe}{\on{Spec}}
\nc{\rv}{\crho}
\nc{\can}{\on{can}}
\nc{\CC}{\on{Op}_G(D))}
\nc{\Op}{\on{Op}_G(D)}
\nc{\MOp}{\on{MOp}_G(D)}
\nc{\Db}{{\mathbb D}}
\nc{\ww}{w}

\nc{\Con}{\on{Conn}(\Omega^{\crho})_D}
\nc{\ConD}{\on{Conn}(\Omega^{\crho})_{\Db}}
\nc{\ConDL}{\on{Conn}(\Omega^{\rho})_{\Db}}
\nc{\ConDtL}{\on{Conn}(\Omega^{\rho})_{\Db^\times}}
\nc{\OpD}{\on{Op}_G(\Db)}
\nc{\crho}{\check{\rho}}
\nc{\chal}{\check{\al}}
\nc{\cchi}{\check{\chi}}
\nc{\cLa}{\check\Lambda}
\nc{\cla}{\check\la}
\nc{\cmu}{\check\mu}

\nc{\PP}{{\mathbb P}}
\nc{\TT}{{\mathbb T}}

\nc{\bone}{{\mb 1}}

\nc{\bs}{\backslash}

\begin{document}

\title{Mirror symmetry in two steps: A--I--B}

\author[Edward Frenkel]{Edward Frenkel$^1$}\thanks{$^1$Partially
  supported by the DARPA grant HR0011-04-1-0031 and the NSF
  grant DMS-0303529}

\address{Department of Mathematics, University of California,
  Berkeley, CA 94720, USA}

\author[Andrei Losev]{Andrei Losev$^2$}\thanks{$^2$Partially
  supported by the Federal Program 40.052.1.1.1112, by the Grants INTAS
  03-51-6346, NSh-1999/2003.2 and RFFI-04-01-00637}

\address{Institute of Theoretical and Experimental Physics,
  B. Cheremushkinskaya 25, Moscow 117259, Russia}

\date{May 2005. Revised December 2005.}

\begin{abstract}

We suggest an interpretation of mirror symmetry for toric varieties
via an equivalence of two conformal field theories. The first theory
is the twisted sigma model of a toric variety in the infinite volume
limit (the A--model). The second theory is an intermediate model,
which we call the I--model. The equivalence between the A--model and
the I--model is achieved by realizing the former as a deformation of a
linear sigma model with a complex torus as the target and then
applying to it a version of the $T$--duality. On the other hand, the
I--model is closely related to the twisted Landau-Ginzburg model (the
B--model) that is mirror dual to the A--model. Thus, the mirror
symmetry is realized in two steps, via the I--model. In particular, we
obtain a natural interpretation of the superpotential of the
Landau-Ginzburg model as the sum of terms corresponding to the
components of a divisor in the toric variety. We also relate the
cohomology of the supercharges of the I--model to the chiral de Rham
complex and the quantum cohomology of the underlying toric variety.

\end{abstract}

\maketitle

\section*{Introduction}

Two-dimensional supersymmetric sigma models have attracted a lot of
attention in recent years.  These models are rich enough to display
many important and non-trivial physical phenomena, understanding which
may help us gain insights into more difficult models, such as the
four-dimensional gauge theories. One of the most interesting phenomena
is mirror symmetry which is a duality between a type A twisted sigma
model and a type B twisted topological theory, such as a
Landau-Ginzburg model (see, e.g., \cite{HV}). The advent of mirror
symmetry has led to spectacular conjectures and results in
mathematics, bringing together such diverse topics as enumerative
algebraic geometry, Gromov-Witten invariants, Floer cohomology,
soliton equations and singularity theory (see the book \cite{Mirror}
and references therein).

In this paper we suggest an interpretation of mirror symmetry for
toric varieties. We show that there is a certain conformal field
theory (the ``I--model'') that is intermediate between the type A
twisted sigma model and the type B twisted Landau-Ginzburg model. On
the one hand, this model is equivalent to the sigma model of a toric
variety in the infinite volume limit, considered as a conformal field
theory, and on the other hand its BPS sector is closely related to the
BPS sector of the corresponding Landau-Ginzburg model. Let us describe
this correspondence in more detail.

\subsection*{Sigma model in the infinite volume}

Consider the type A twisted $N=(2,2)$ supersymmetric sigma model with
a target K\"ahler manifold $M$. This model is believed to define a
superconformal quantum field theory if the K\"ahler metric is Ricci
flat, i.e., if $M$ is a Calabi-Yau manifold. However, we will argue in
this paper that a suitable {\em infinite volume} limit of the twisted
sigma model defines a conformal field theory for more general target
manifolds.

This infinite volume limit is defined at the classical level by
passing to a suitable ``first order formalism'' Lagrangian, which has
previously been considered in the literature in \cite{W:tsm,BS,Moore}
and more recently in \cite{BLN,Kapustin}. Rescaling the K\"ahler
metric by a parameter $t$, we find that the first order Lagrangian has
a well-defined limit even as $t \to \infty$. In this limit we obtain a
conformally invariant Lagrangian, which describes what is natural to
call the infinite volume limit of the twisted sigma model (see
\secref{first order} for details).

Quantization of a first order Lagrangian could be non-trivial and even
problematic in some cases. However, in the twisted $N=(2,2)$
supersymmeric theory that we are considering it is expected that all
potential anomalies cancel out and the theory remains conformally
invariant at the quantum level as well. Moreover, the corresponding
path integral over all maps $\Phi: \Sigma \to M$, where $\Sigma$ is a
Riemann surface (the worldsheet), has a nice geometric interpretation
as the delta-form supported on the subspace of {\em holomorphic maps}
$\Phi: \Sigma \to M$. When we deform the Lagrangian back to the finite
volume, i.e., to finite values of $t$, we obtain what looks like a
``smoothening'' of this delta-form, or, more precisely, the
Mathai-Quillen representative of the Euler class of an appropriate
vector bundle over the space of maps, see \cite{Moore}. Hence it is
natural to think that in the infinite volume limit the path integral
localizes on the holomorphic maps, i.e., it can be represented as a
sum of integrals over the finite-dimensional moduli spaces of
holomorphic maps of different degrees (see
\secref{quantization}). This is what one expects in the type A twisted
sigma model in the infinite volume limit as explained by E. Witten in
\cite{W:sdg,W:mirror}.

We wish to view the model in the infinite volume first and foremost as
a topological conformal field theory. In particular, it should come
with a Hilbert space combining chiral and anti-chiral states, and a
state-field correspondence. Correlation functions should be defined
for any Riemann surface $\Sigma$ with marked points $x_1,\ldots,x_n$
(and possibly germs of local coordinates at those points), and a
collection of local operators inserted at those points. These
correlation functions may be viewed as differential forms on the
moduli space ${\mathcal M}_{g,n}$ of pointed curves $(\Sigma,(x_i))$
(see \secref{cor fn}).

Part of this structure is captured by the {\em Gromov-Witten
invariants}, which appear as integrals of the differential forms
corresponding to particular observables over a compactification of
${\mathcal M}_{g,n}$ (see \secref{cor fn} for more details).

Another ingredient of this conformal field theory is a sheaf of chiral
algebras over $M$, called the {\em chiral de Rham complex}, introduced
in \cite{MSV}. It is defined by gluing free chiral algebras on the
overlaps of open subsets of $M$ isomorphic to $\C^n$. From the point
of view of the twisted sigma model, this chiral algebra corresponds to
the cohomology of the right moving supercharge in the {\em
perturbative} regime, i.e., without counting the instanton
contributions, as explained in \cite{W:new,Kapustin}. In order to
understand the correlation functions of the sigma model and in
particular to include the instanton corrections, it is necessary to go
beyond the chiral algebra and consider the full conformal field
theory. This is one of the goals of the present paper.

\subsection*{Non-linear sigma models as deformations of free field
  theories}

There is one case when the sigma model can certainly be defined as a
conformal field theory, and this is the case of the target manifolds
with a flat metric, such as a flat space $\C^n$ or a torus (for a
detailed treatment of the latter, see \cite{KO}). We will consider in
\secref{toric sigma model} the intermediate case of the sigma model in
the infinite volume with the target manifold a complex torus
$(\C^\times)^n$, which we call the {\em toric sigma model}. This is a
free conformal field theory, but we will show that it exhibits some
non-trivial effects, such as the appearance of holomorphic analogues
of vortex operators, which we call {\em holomortex} operators.

We will then define in Sections \ref{changing to pone} and
\ref{general toric} the conformal field theory governing a non-linear
sigma model of a toric variety in the infinite volume as a {\em
deformation}, in the sense of A. Zamolodchikov \cite{Zam}, of the
toric sigma model, by some explicitly written exactly marginal
operators. By its very definition, this deformed conformal field
theory will include the instanton effects corresponding to holomorphic
maps of non-zero degree.

To illustrate our main idea, it is instructive to look at the case of
the sigma model with the target $\pone$ in the infinite volume limit,
obtained by quantization of the corresponding first order Lagrangian.
We wish to obtain it as a deformation of the toric sigma model with
the target $\C^\times$ which we realize as the quotient $\C/2\pi i
\Z$. This is a free conformal field theory with the basic chiral
fields $X(z), p(z)$, $\psi(z)$, $\pi(z)$, and their anti-chiral
partners with the action
\begin{equation}    \label{action first time}
\frac{i}{2\pi} \int_{\Sigma} d^2 z \; \left( p \pa_{\ol{z}} X + \ol{p}
\pa_z \ol{X} + \pi \pa_{\ol{z}} \psi + \ol\pi \pa_z \ol\psi \right).
\end{equation}
The field $X(z)$ corresponds to a linear coordinate on $\C/2\pi i \Z$,
and so is defined modulo $2\pi i \Z$.

As discussed above, the correlation functions of this model are given
by integrals over the space of holomorphic maps $\Sigma \to
\C^\times$. For compact $\Sigma$, all such maps are necessarily
constant. Therefore the correlation functions reduce to integrals over
the zero mode (i.e., over the image of the constant map $\Phi: \Sigma
\to \C^\times$), as expected in a free field theory.

How can we interpret holomorphic maps $\Sigma \to \pone$ within the
framework of this free field theory? Such maps may be viewed as
holomorphic maps $\Sigma \bs \{ w_i^{\pm} \} \to \C/2\pi i \Z$ with
logarithmic singularities at some points $w^\pm_1,\ldots,w^\pm_N$,
where this map behaves as $\pm \log(z-w^\pm_i)$. These singular points
correspond to zeroes and poles of $\exp \Phi$, and generically they
will be distinct. Our proposal is that {\em we can create these
singularities of $\Phi$ by inserting in the correlation function of
the linear sigma model certain vertex operators} $\Psi_\pm(w^\pm_i)$.

The defining property of the operators $\Psi_\pm(w)$ (up to a scalar)
is that their operator product expansion (OPE) with $X(z)$ should read
\begin{equation}    \label{property}
X(z) \Psi_\pm(w) = \pm \log(z-w) \Psi_\pm(w).
\end{equation}
Given such operators, we can write a given function (in the case
of $\Sigma$ of genus zero)
$$
\Phi(z) = c + \sum_{i=1}^n \log(z-w^+_i) - \sum_{i=1}^n \log(z-w^-_i)
$$
as the correlator
$$
\Phi(z) = \langle X(z) \prod_{i=1}^n \Psi_+(w^+_i) \prod_{i=1}^n
\Psi_-(w^-_i) \delta^2(X(\infty)-c) \psi(\infty) \ol\psi(\infty) \rangle
$$
(the term involving the delta-function and the fermions will give,
upon the integration over the zero modes of $X$ and $\psi$, the
normalization condition $\Phi(\infty) = c$). Thus, we can create all
instantons of the $\pone$ sigma model, that is holomorphic maps
$\Sigma \to \pone$, as correlation functions in the toric sigma model
of the above form (the case of $\Sigma$ of genus greater than zero
will be discussed in \secref{deformation}).

The property \eqref{property} is satisfied by the following fields
$$
\Psi_\pm(w,\ol{w}) = \exp \left( \mp i \int_{w_0}^w (p(z) dz +
\ol{p}(\ol{z}) d\ol{z}) \right),
$$
which are examples of the holomortex operators mentioned
above. Including these operators in the correlation functions and
allowing the points $w^\pm_i$ to vary over $\Sigma$ is equivalent to
deforming the action \eqref{action first time} with the term
$$
q^{1/2} \int_\Sigma \left( \Psi^{(2)}_+ + \Psi^{(2)}_- \right),
$$
where $\Psi^{(2)}_\pm$ are the cohomological descendants
$$
\Psi^{(2)}_\pm = \Psi_\pm(w,\ol{w}) \pi(w) \ol{\pi}(\ol{w}) dw d\ol{w}.
$$
The resulting deformed theory appears to be equivalent to the sigma
model with the target $\pone$ in the infinite volume limit (in the
sense explained in \secref{changing to pone}). By construction, the
part of a correlation function of this deformed theory that
corresponds to degree $n$ maps $\Sigma \to \pone$ will appear with the
overall factor $q^n$.

More generally, suppose that we are given a smooth compact K\"ahler
manifold $M$ with an open dense submanifold $M_0$ with a linear
structure. The complement $C = M\backslash M_0$ is a compactification
divisor, which is a union of irreducible components
$C_1,\ldots,C_N$. The linear sigma model corresponding to $M_0$ is a
free superconformal field theory, and we wish to describe the
non-linear model with the target $M$ in terms of this theory. Let us
observe that a generic holomorphic map $\Phi: \Sigma \to M$ will take
values in $C$ at a finite set of points $x_1,\ldots,x_n$, and
generically we will have $\Phi(x_j) \in C_{k_j}$ and $\Phi(x_j)
\not\in C_l, l \neq k_j$. To account for such maps we need to insert
some vertex operators $\Psi_{k_j}$ corresponding to the
compactification divisors $C_{k_j}$ at the points $x_j,
j=1,\ldots,n$. It is then natural to expect that the non-linear sigma
model with the target $M$ in the infinite volume limit can be
described as the deformation of the free field theory corresponding to
the target $M_0$ by means of the operators $\Psi^{(2)}_k,
k=1,\ldots,N$, where $\Psi^{(2)}_k$ is the $(1,1)$--form counterpart
of $\Psi_k$ obtained via the cohomological descent. To solve the
theory we therefore need to identify explicitly the suitable vertex
operators $\Psi_k, k=1,\ldots,n$, corresponding to the
compactification divisors. In general, they may be highly non-local
and given by very complicated formulas.

While finding these operators may seem like a daunting task in
general, it turns out that in the case when the target is a toric
variety, they can be written down quite explicitly. Such a variety $M$
comes with a natural open dense subset $M_0$ isomorphic to
$(\C^\times)^n$ and the compactification divisors are naturally
parameterized by the one-dimensional cones in the fan defining $M$. We
construct explicitly the vertex operators corresponding to these
compactification divisors in \secref{general toric}. These operators
may be viewed as holomorphic counterparts of the vortex operators
familiar from the free bosonic theory compactified on a torus. We will
argue that the deformation of the action by these operators changes
the topology of the target manifold and deforms a free field theory to
a non-linear sigma model with the target $M$.

As in the case of $\pone$, we expect that the sigma model with the
target $M$, which is a smooth compact Fano toric variety, is
equivalent to a deformation of the free field theory with the target
$(\C^\times)^n$ by the holomortex operators corresponding to the
irreducible components of the compactification divisor.

As a consistency check, we compute in \secref{cohomology} the
cohomology of the right moving supercharge in our deformed theory,
making a connection to the results of L. Borisov \cite{Bor} and
F. Malikov--V. Schechtman \cite{MS}. In particular, we show that in
the case of $M=\PP^n$ this cohomology is equal to the quantum
cohomology of $\PP^n$. On the other hand, in a certain limit we obtain
the cohomology of the chiral de Rham complex of $M$. This is
consistent with the assertion of \cite{W:new,Kapustin} that the chiral
de Rham complex should appear as the cohomology of the right moving
supercharge of the type A twisted sigma model in the perturbative
regime.

\subsection*{I--model and mirror symmetry}

Next, we consider the question as to what is the meaning of {\em
mirror symmetry} from the point of view of our description of the
sigma model of a toric variety as a deformation of a free field
theory. The first step in answering this question is to perform a kind
of $T$--duality transform of the free field theory with the target
$(\C^\times)^n$.

In the case of $\pone$, before the deformation, we have the free field
theory with the target $\C^\times$. The dual of this theory turns out
to be the ordinary sigma model with the target being the cylinder $\R
\times S^1$ equipped with the metric of {\em Minkowski} signature. Let
$R$ and $U$ be the coordinates on $\R$ and $S^1 = \R/2\pi$,
respectively. Under the $T$--duality the local fields $p$ and $X$
become more complicated, but the complicated fields, like the
holomortex operators, become simple. In fact, we have the following
transformation:
$$
p dz + \ol{p} d\ol{z} = d U,
$$
and so the holomortex operators $\Psi_\pm$ turn out to be simply the
exponential fields $e^{\mp i U}$.  The field $R$ coincides with the
field $\frac{1}{2}(X+\ol{X})$ of the original theory. Therefore $e^R$
coincides with the field $|e^X|$, the absolute value of the
holomorphic coordinate on $\pone$ compactifying the target
$\C^\times$. The action of the deformed dual theory reads
\begin{equation}    \label{i model}
\frac{i}{2\pi} \int_{\Sigma} d^2 z \; \left( \pa_z U \pa_{\ol{z}} R
+ \pa_{\ol{z}} U \pa_{z} R + \pi \pa_{\ol{z}} \psi + \ol\pi
\pa_z \ol\psi \right) + q^{1/2} \int_\Sigma (e^{iU} + e^{-iU}) \pi
\ol{\pi} d^2 z.
\end{equation}
Thus, the correlation functions of the observables of this theory that
depend only on the field $R$ realize the corresponding correlation
functions of the twisted sigma model, namely, those that depend only
on $|e^X|$. But while the correlation functions of the twisted sigma
model appear as sums over the instanton contributions, the dual
description gives us their {\em non-perturbative} realization!

Let us compare the action \eqref{i model} to the action of the
Landau-Ginzburg model with the target $\C$ and the Landau-Ginzburg
superpotential $$W = q^{1/2}(e^{iY} + e^{-iY}),$$ where $Y$ is a
chiral superfield:
\begin{multline}    \label{LG}
\frac{1}{2\pi} \int_\Sigma d^2 z \; \left( \pa_z \varphi \pa_{\ol{z}}
\ol{\varphi} + \pa_{\ol{z}} \varphi \pa_{z} \ol{\varphi} + i \chi_+
\pa_{\ol{z}} \ol\chi_+ + i \chi_- \pa_z \ol\chi_- \right) \\ + q^{1/2}
\int_\Sigma (e^{i\varphi} + e^{-i\varphi}) \chi_+ \chi_- d^2 z.
\end{multline}
We observe that the two actions look similar: if we ``analytically
continue'' the theory with the action \eqref{i model}, allowing the
fields $U$ and $R$ to become complex-valued fields $\varphi$ and
$\ol{\varphi}$, which are complex conjugate to each other, and rename
the fermions as follows: $$\pi \mapsto \chi_-, \quad \ol\pi \mapsto
\chi_+, \quad \psi \mapsto \ol\chi_-, \quad \ol\psi \mapsto
\ol\chi_+,$$ then the action \eqref{i model} becomes the action
\eqref{LG}. This means that the correlation functions in the two
theories should be related by a kind of analytic
continuation. However, we wish to stress the models with the actions
\eqref{i model} and \eqref{LG} are {\em different}. For example, in
the model \eqref{i model} the field $U$ is real periodic, and $R$ is
real non-periodic, while in the model \eqref{LG} the fields $\varphi$
and $\ol\varphi$ are complex (conjugate to each other) and both
periodic.

It is instructive to compare the supersymmetry charges in the above
models. For simplicity we consider the case when $q=0$. In the
original A--model with the action \eqref{action first time} the
supercharge is
$$
\int (\psi p dz + \ol\psi \ol{p} d\ol{z}).
$$
This is a de Rham type supercharge, because under its action $X
\mapsto \psi, \ol{X} \mapsto \ol\psi$. In the $T$--dual theory with
the action \eqref{i model} (with $q=0$) the supercharge becomes
$$
\int (\psi \pa_z U dz + \ol\psi \pa_{\ol{z}} U d\ol{z}).
$$
Under the ``analytic continuation'' that we discussed above, it
becomes the supercharge of the type B twisted Landau-Ginzburg model
with the action \eqref{LG} (with $q=0$):
$$
\int (\ol\chi_- \pa_z \varphi dz + \ol\chi_+ \pa_{\ol{z}} \varphi
d\ol{z}).
$$
This is now a Dolbeault type supercharge, because under its action
$\varphi \mapsto 0, \ol\varphi \mapsto \ol\chi_- + \ol\chi_+$. Thus,
the $T$--duality indeed transforms a de Rham type supercharge of the
A-model to a Dolbeault type supercharge of the B--model, as expected
in mirror symmetry.  Note that the interpretation of the fermionic
fields is very different in the two theories, and this underscores the
highly non-local nature of the mirror symmetry.

Traditionally the Landau-Ginzburg model is defined by adding to the
action of the supersymmetric linear sigma model the term $\int d^2 z
d^2 \theta W(Y) + \int d^2 z d^2 \ol\theta \,
\ol{W}(\ol{Y})$. Usually, one chooses $\ol{W}(\ol{Y})$ to be complex
conjugate of $W(Y)$. But in a type B twisted Landau-Ginzburg model
there is an essential difference between the first and the second
terms: while the integrand in the first one is a $(1,1)$--form, the
integrand in the second is a $(0,0)$--form, and hence to integrate it
one needs to pick a metric on the worldsheet. This breaks conformal
invariance. That is why in the action \eqref{LG} we have set
$\ol{W}=0$, for otherwise the theory would not be conformally
invariant.

The Landau-Ginzburg model with the (twisted) superpotential $W$, where
$W$ is as above, and its complex conjugate $\ol{W}$ has been
considered by K. Hori and C. Vafa \cite{HV} (see also
\cite{FI,CV,EHY,Giv}). They showed that its correlation functions {\em
in the BPS sector} are related to those of the twisted sigma model of
$\pone$, which is the sense in which the two theories are mirror dual
to each other. Note that $\ol{W}$ is $Q$--exact, and the possibility
of setting $\ol{W}$ to $0$ was mentioned in \cite{Losev} and
\cite{HV}, Sect. 6.

The point of our construction is that in addition to the twisted sigma
model and the Landau-Ginzburg model, which are usually considered in
the study of mirror symmetry, there is an intermediate model, or the
``I--model'', described by the action \eqref{i model}. This is a
conformal field theory that has two properties: on the one hand it
should be equivalent to the type A twisted sigma model with the target
$\pone$ in the infinite volume, which is also a conformal field
theory. In other words, {\em all} correlation functions in the two
models are equivalent, not just in the BPS sector. On the other hand,
the BPS sector of the I--model is closely related to the BPS sector of
the type B twisted Landau-Ginzburg model considered in \cite{HV} (see
the discussion in \secref{dual} for more details).

This conclusion leads to a curious observation that the correlation
functions of the field $e^R$ in the I--model (which corresponds to
$e^{\ol{\varphi}}$ in the Landau-Ginzburg model \eqref{LG}) encode the
correlation functions of the field $|e^X|$ of the sigma model with the
target $\pone$. Thus, one can actually {\em see the $\pone$
instantons}, and not just the correlation functions of the BPS states,
in the framework of the I--model (or the Landau-Ginzburg model)!

We define a similar I--model for an arbitrary toric variety. Then the
corresponding deformation term in the Lagrangian is equal to the sum
$\sum_{k=1}^N e^{- iU_k} \pi_{(k)} \ol\pi_{(k)}$ over the
components of the compactification divisor of our toric variety. The fields
$U_k$ satisfy constraints reflecting the structure of the fan defining
the toric variety $M$. For example, in the case when $M={\mathbb P}^n$
we have $N=n+1$, and the fields $U_k$ satisfy the familiar constraint
$\prod_{k=1}^{n+1} e^{- iU_k} = q$. Thus, we immediately recognize
that, after the analytic continuation, we obtain a term that looks
like the Landau-Ginzburg superpotential corresponding to $\PP^n$
considered in \cite{HV}. We note that these superpotentials and the
corresponding oscillating integrals representing correlation functions
of the Landau-Ginzburg model had previously appeared in the
mathematical work of A. Givental \cite{Giv} on mirror symmetry.

We stress that in our approach the superpotential is generated because
of our description of the sigma model with the target $M$ (in the
infinite volume limit) as a deformation of a free field theory, to
which we apply the $T$--duality transform. Therefore the
superpotential has a transparent geometric meaning. Namely, the
summands appearing in the superpotential naturally correspond to the
irreducible components of the compactification divisor in $M$. The mirror
symmetry can now be viewed as a corollary of the equivalence of the
I--model and the A--model (sigma model with the target $M$ in the
infinite volume limit), as conformal field theories. We hope that the
I--model will help us understand more fully the phenomenon of mirror
symmetry.\footnote{It is instructive to compare our derivation of
mirror symmetry to A. Polyakov's model of confinement in three
dimensions \cite{Pol}.}

In the case of $\PP^n$, the action of the I--model is very similar to
the action of the $A_{n-1}^{(1)}$ affine Toda field theory, considered
as a deformation of a free field theory. However, since the I--model
is conformally invariant, its structure is actually more reminiscent
of that of the conformal $A_{n-1}$ Toda field theory. We can use the
methods familiar from the Toda theory to determine the structure of
the chiral sector of the I--model. We recall that in the case of an
$A_{n-1}$ Toda field theory the chiral algebra of integrals of motion
is the $\W_n$--algebra \cite{FL,FF:laws}. It appears as the subalgebra
of those operators of the free field theory which commute with the
{\em screening operators}, which are the residues of the operators
deforming the action. Likewise, the $\W$--algebra in the I--model
associated to a toric variety $M$ consists of the operators that
commute with the operators $\int e^{- iU_k} \pi_{(k)} dz,
k=1,\ldots,n+1$ (which can therefore be viewed as supersymmetric
analogues of the screening operators), and it is possible to determine
it explicitly. In doing so, we make a connection to the results of
\cite{Bor} (see also \cite{F,GMS,MS}) and show that this $\W$--algebra
is isomorphic to the algebra of global sections of the chiral de Rham
complex on $M$.

In a follow-up paper we will generalize our results to hypersurfaces
in toric varieties, and, more generally, to complete intersections in
toric varieties. This way we hope to obtain a realization of mirror
symmetry for such varieties as an equivalence of conformal field
theories in the sense explained above.

In a future work we plan to consider an analogue of this construction
for the $(0,2)$ supersymmetric sigma models. We believe that in the
case when $M$ is a flag manifold of a simple Lie group, this theory,
when coupled to gauge theory, is closely related to the geometric
Langlands correspondence. We also plan to apply similar methods to the
study of four-dimensional supersymmetric Yang-Mills theories.

The paper is organized as follows. In \secref{limit} we discuss the
sigma model in the infinite volume limit, at both classical and
quantum levels. We explain how the first order Lagrangian (with a
B--field term) arises in the infinite volume limit and the
interpretation of the corresponding path integrals as integrals of
differential forms on the moduli spaces of holomorphic maps. We then
outline our idea of constructing non-linear sigma models as
deformations of linear ones. We illustrate this idea on the example of
the deformation of the target manifold from $\C$ to $\pone$. In
\secref{toric sigma model} we introduce the toric sigma model, which
is the linear sigma model with the target $\C^\times$ in the infinite
volume. We define the holomortex operators and the $T$--duality
transform. We show that the $T$--dual model of the toric sigma model
is the ordinary sigma model with the target being the cylinder
equipped with a metric of Minkowski signature. In \secref{changing to
pone} we consider a deformation of the toric sigma model to the sigma
model with the target $\pone$. We then define the T--dual theory,
which is our I--model. We give a sample computation of the correlation
functions in the I--model and obtain explicit formulas for the
supercharges. We generalize these results to the case of an arbitrary
compact smooth toric variety in \secref{general toric}. Finally, we
discuss the operator formalism of these theories in \secref{operator},
as well as their $\W$--algebras and the cohomologies of the
supersymmetry charges.

\subsection*{Acknowledgments}

E.F. is grateful to E. Witten for illuminating discussions of sigma
models, and in particular for explaining his ideas on the connection
between sigma models and the chiral de Rham complex prior to their
publication in \cite{W:new}. E.F. also thanks B. Feigin, A.J. Tolland,
and especially A. Givental for useful discussions.

A.L. thanks B. Feigin, M. Olshanetsky and A. Rosly for discussions.

We are grateful to N. Nekrasov and E. Witten for their comments on
a draft of this paper.

Our collaboration on this project started during a meeting on the
geometric Langlands correspondence in Chicago in October of 2004 that
was sponsored by DARPA through its Program ``Fundamental Advances in
Theoretical Mathematics''. We gratefully acknowledge support from
DARPA.

\section{Supersymmetric sigma model in the infinite volume limit}
\label{limit}

\subsection{Lagrangian description}    \label{first order}

We start by describing the A twisted $N=(2,2)$ supersymmetric sigma
model in the formalism of the first order, following
\cite{W:tsm,BS,Moore}. Let $\Sigma$ be a complex Riemann surface
(worldsheet). We denote by $z$ and $\ol{z}$ the local holomorphic and
anti-holomorphic coordinates on $\Sigma$, and by $d^2 z = i dz \wedge
d \ol{z}$ the corresponding integration measure on $\Sigma$. Let $M$
be a complex K\"ahler manifold (target) with a fixed K\"ahler metric
$g_{a\ol{b}}$.

We will denote by $X^a, a=1,\ldots,N=\dim M$, local holomorphic
coordinates on $M$, and by $X^{\ol{a}} = \ol{X^a}$ their complex
conjugates. Given a map $\Phi: \Sigma \to M$, we consider the
pull-backs of $X^a$ and $X^{\ol{a}}$ as functions on $\Sigma$, denoted
by the same symbols. We also have fermionic fields $\psi^a$ and
$\psi^{\ol{a}}, a=1,\ldots,N$, which are sections of $\Phi^*(T^{1,0}
M)$ and $\Phi^*(T^{0,1} M)$, respectively. The Levi-Civita connection
on $T M$ corresponding to the metric $g_{a\ol{b}}$ induces a
connection on $\Phi^*(T M)$. The corresponding covariant derivatives
have the form
\begin{align*}
D_{\ol{z}} \psi^a &= \pa_{\ol{z}} \psi^a + \pa_{\ol{z}}
X^b \cdot \Gamma^a_{bc} \psi^c, \\
D_z \psi^{\ol{a}} &= \pa_z \psi^{\ol{a}} +  \pa_z X^{\ol{b}} \cdot
\Gamma^{\ol{a}}_{\ol{b}\ol{c}} \psi^{\ol{c}},
\end{align*}
where $\Gamma^a_{bc} = g^{a \ol{b}} \pa_{b} g_{c \ol{b}}$.

Next, we introduce auxiliary fields $p_a$ which will play the role of
the ``Lagrange multipliers'' corresponding to the equations
$\pa_{\ol{z}} X^a = 0$, and their complex conjugates
$p_{\ol{a}}$. Their fermionic super-partners will be denoted by
$\pi_a$ and $\pi_{\ol{a}}$. These are sections of $\Phi^*(\Omega^{1,0}
M) \otimes \Omega^{1,0} \Sigma$ and $\Phi^*(\Omega^{0,1} M) \otimes
\Omega^{0,1} \Sigma$, respectively.

We write down the action for these fields following \cite{W:tsm}
(formula (2.14)):
\begin{multline}    \label{action}
I_t = \frac{1}{2\pi} \int_{\Sigma} d^2 z \; \left( i p_a \pa_{\ol{z}}
X^{a} + i p_{\ol{a}} \pa_z X^{\ol{a}} + i \pi_a D_{\ol{z}} \psi^{a} +
i \pi_{\ol{a}} D_{z} \psi^{\ol{a}} \right. \\ - \left. t^{-1}
R^{a\ol{b}}{}_{c\ol{d}} \pi_a \pi_{\ol{b}} \psi^c \psi^{\ol{d}} +
t^{-1} g^{a\ol{b}} p_a p_{\ol{b}} \right),
\end{multline}
where $t$ is a parameter (the ``radius''). The equations of motion for
$p_a, p_{\ol{a}}$ are as follows:
\begin{align}    \label{eqs of motion1}
p_a &= -i t g_{a\ol{b}} \pa_z X^{\ol{b}}, \\
p_{\ol{a}} &= -i t g_{\ol{a}b} \pa_{\ol{z}} X^b. \notag
\end{align}

\begin{remark}    \label{misleading}
Formulas \eqref{eqs of motion1} seem to indicate that the complex
conjugate of $p^{a}$ is equal to $-p^{\ol{a}}$, which is
misleading. In fact, the substitution \eqref{eqs of motion1} is formal
and only makes sense under the path integral. It corresponds to
completing the action to a square and integrating out the variables
$p^a$ and $p^{\ol{a}}$.\qed
\end{remark}

Substituting these formulas back into \eqref{action}, we obtain the
action
$$
\wt{I}_t = \frac{1}{2\pi} \int_{\Sigma} d^2 z \; \left( t g_{a\ol{b}}
\pa_{\ol{z}} X^a \pa_z X^{\ol{b}} + i \pi_a D_{\ol{z}} \psi^{a} + i
\pi_{\ol{a}} D_{z} \psi^{\ol{a}} - t^{-1} R^{a\ol{b}}{}_{c\ol{d}}
\pi_a \pi_{\ol{b}} \psi^c \psi^{\ol{d}} \right).
$$
This is the action of the A--twisted $N=(2,2)$ supersymmetric sigma
model with the target $M$ and the B--field $- \frac{t}{2\pi}
\omega$, where $\omega = \frac{i}{2} g_{a\ol{b}}
dX^a \wedge dX^{\ol{b}}$ is the K\"ahler form on $M$, introduced in
\cite{W:tsm,W:mirror}. The corresponding metric on $M$ is $t
g_{a\ol{b}}$. Thus, the action \eqref{action} describes this model. In
the infinite volume limit $t \to \infty$ the action \eqref{action}
becomes
\begin{equation}    \label{action infty}
I_\infty = \frac{i}{2\pi} \int_{\Sigma} d^2 z \; \left( p_a
\pa_{\ol{z}} X^{a} + p_{\ol{a}} \pa_z X^{\ol{a}} + \pi_a D_{\ol{z}}
\psi^{a} + \pi_{\ol{a}} D_{z} \psi^{\ol{a}} \right).
\end{equation}
This action is conformally invariant, and it has two supersymmetries:
one is mapping
\begin{align*}
X^a &\mapsto \psi^a, \qquad \psi^a \mapsto 0, \\
\pi_a &\mapsto - p_a - \Gamma^b_{ac} \pi_b \psi^c, \\
p_a &\mapsto \Gamma^b_{ac} p_b \psi^c,
\end{align*}
and the other does the same to their complex conjugates.

\subsection{The path integral}    \label{quantization}

The action \eqref{action infty} describes a conformal field theory
governing the infinite volume limit of the A--twisted sigma model. We
wish to understand the corresponding quantum field theory.

The first observation is that the path integral $\int [Dp] [D\pi]
e^{-I_\infty}$, considered as a differential form on the space of maps
$\Sigma \to M$, may be viewed as the integral representation of the
delta-function differential form supported on the space of holomorphic
maps $\Sigma \to M$.

To see this, consider a finite-dimensional model situation: a complex
vector space $\C^M$ and functions $f^a, a=1,\ldots,N$, defining a
codimension $N$ complex subvariety $C \subset \C^M$. Then the
delta-like differential form supported on this subvariety has the
following integral representation:
$$
\delta_C = \int \prod_a dp_a dp_{\ol{a}} d \pi_a d\pi_{\ol{a}} \exp
\left( - i p_a f^a - i \ol{p}_a \ol{f}_a - i \pi_a df_a - i \ol\pi_a
d\ol{f}_a \right).
$$
This delta-form may be viewed as the limit, when $t \to \infty$, of
the regularized integral
$$
\delta_{C,t} = \int \prod_a dp_a dp_{\ol{a}} d \pi_a d\pi_{\ol{a}}
  \exp \left( - i p_a f^a - i \ol{p}_a \ol{f}_a - i \pi_a df_a - i
\ol\pi_a d\ol{f}_a - t^{-1} p_a p_{\ol{a}} \right).
$$
Comparing these formulas to \eqref{action} and \eqref{action infty},
we see that the path integral
\begin{equation}    \label{path}
\int [Dp] [D\pi] e^{-I_\infty}
\end{equation}
looks like a delta-like form supported on the solutions of the
equation $\pa_{\ol{z}} X^{a} = 0$, i.e., on the holomorphic maps,
while $\int [Dp] [D\pi] e^{-I_t}$ may be viewed as its regularized
version. Alternatively, and more precisely, one may say that the
integral $\int [Dp] [D\pi] e^{-I_t}$ looks like the Mathai-Quillen
representative of the Euler class of an appropriate vector bundle over
the space of maps $\Sigma \to M$ (see \cite{Moore}, \S~13.6).

Motivated by this analogy, it is natural to expect that in the
infinite volume limit the correlation functions in our theory will
correspond to sums of integrals of differential forms over different
connected components of the moduli space of holomorphic maps $\Sigma
\to M$, as explained in \cite{W:sdg}. Particular examples of these
functions give rise to the Gromov-Witten invariants of $M$
\cite{W:mirror}.

The connected components of the moduli space of holomorphic maps
$\Sigma \to M$ are labeled by $H_2(M)$. Choosing a basis in $H_2(M)$,
we can label them by $k$--tuples of integers $(n_1,\ldots,n_k)$. It is
customary to weight the contribution to the path integral
corresponding to the component of the space of holomorphic maps
$\Sigma \to M$ of degree $(n_1,\ldots,n_k)$ with the coefficient
$q_1^{n_1} \ldots q_k^{n_k}$ (we choose this basis in such a way that
non-zero contributions come from $n_i \geq 0$). This can be achieved
by adding to the action $I_\infty$ the topological term $\sum_i
\frac{u_i}{2\pi} \int_{\Sigma} \Phi^*(\varpi^i)$. Here $\{ \varpi^i
\}$ is the basis of the K\"ahler cone of $M$ that is dual to the above
basis of $H_2(M)$ and the $u_i$'s are the coupling constants such that
$q_i=e^{-u_i}$.

The corresponding path integral is then the sum over $n_1,\ldots,n_k
\geq 0$ of terms corresponding to the holomorphic maps $\Sigma \to M$
of degrees $(n_1,\ldots,n_k)$ with coefficients $q_1^{n_1} \ldots
q_k^{n_k}$. This path integral may be obtained as the $t \to \infty$
limit of a sigma model path integral as follows. We simply add
to the action the B--field $-\frac{t}{2\pi} \omega + \frac{1}{2\pi}
\varpi$, where $\omega = \frac{i}{2} g_{a\ol{b}} dX^a \wedge
dX^{\ol{b}}$ is the K\"ahler form on $M$ and $\varpi = \sum_i u_i
\varpi^i$. Then the bosonic part of the action will read
$$
\frac{1}{2\pi} \int_\Sigma d^2 z \; \left( \frac{t}{2} \left( g_{a\ol{b}}
\pa_{\ol{z}} X^a \pa_z X^{\ol{b}} + g_{a\ol{b}}
\pa_z X^a \pa_{\ol{z}} X^{\ol{b}} \right) + \frac{t}{2} \left(
g_{a\ol{b}} \pa_{\ol{z}} X^a \pa_z X^{\ol{b}} - g_{a\ol{b}}
\pa_z X^a \pa_{\ol{z}} X^{\ol{b}} \right) \right.
$$
$$
\left. + \sum_i u_i \Phi^*(\varpi^i) \right)
= \frac{1}{2\pi} \int_\Sigma  d^2 z \; t g_{a\ol{b}}
\pa_{\ol{z}} X^a \pa_z X^{\ol{b}} + \sum_i \frac{u_i}{2\pi}
\int_\Sigma \Phi^*(\varpi^i).
$$
In terms of the first order variables this becomes
$$
\frac{1}{2\pi} \int_{\Sigma} d^2 z \; \left( i p_a \pa_{\ol{z}}
X^{a} + i p_{\ol{a}} \pa_z X^{\ol{a}} +
t^{-1} g^{a\ol{b}} p_a p_{\ol{b}} \right) + \sum_i \frac{u_i}{2\pi}
\int_\Sigma \Phi^*(\varpi^i).
$$
Therefore in the limit $t \to \infty$ the path integral will indeed
give us the desired sum over $n_1,\ldots,n_k \geq 0$ weighted with
coefficients $q_1^{n_1} \ldots q_k^{n_k}$, where $q_i = e^{-u_i}$.

Proper definition of the path integral \eqref{path} for worldsheets
$\Sigma$ of genus greater than zero requires a prescription for the
integration of the zero modes of the fields $p_a$ and
$p_{\ol{a}}$. \footnote{We thank N. Nekrasov for a discussion of this
point.} The most evident possibility to do so is to add the term of
the form $\epsilon G^{a\ol{b}} p_a p_{\ol{b}}$ to the action and
consider the limit $\epsilon \to 0$. However, if we choose
$G^{a\ol{b}}$ to be the inverse of a K\"ahler form on $M$, this will
bring us back to the finite volume and spoil conformal invariance if
$M$ is not Calabi-Yau. But we can take $G^{a\ol{b}}$ to be any tensor
in $T^{1,1} M$ of the following form. Suppose that we have a flat
K\"ahler metric on an open dense subset $M_0$ of $M$, such that its
inverse is a section of $T^{1,1} M_0$ that extends to a section on the
entire $M$. We can then take this extension as our $G^{a\ol{b}}$. Then
we can regularize the integrals over the zero modes of the $p_a$'s and
$p_{\ol{a}}$'s without violating conformal invariance of the
theory. Such tensors can be easily constructed for Fano toric
varieties, and we will see examples of that below. We also remark that
for general Fano manifolds the zero modes disappear altogether when
the genus of $\Sigma$ is fixed and the degree of the map $\Sigma \to
M$ is sufficiently high.

\begin{remark}

The action \eqref{action infty} is conformally invariant, and we
expect that the corresponding quantum field theory is also conformally
invariant, for any K\"ahler manifold $M$. However, in the case of
non-Ricci flat Kahler manifolds non-zero $\beta$--function is
developed and the theory becomes non-conformal for finite values of
$t$, even though the deformation to finite volume is achieved by
adding the operator $V = \sum _{a,\ol{b}} g^{a\ol{b}} p_a p_{\ol{b}}$
of dimension (1,1). In general, consider the basis $V_a$ in the space
of operators of dimension (1,1). The $|z-w|^{-2}$ term in their
operator product expansion reads as follows
$$
V_a(z) V_b(w) \sim \frac{C_{ab}^{c} V_c(w)}{|z-w|^2}.
$$
Then the theory with interaction $t^a V_a$ has the beta-function equal
to $t^a t^b C_{ab}^{c} V_c$. In our case, the OPE of the above
operator $V$ with itself contains $|z-w|^{-2}$ with the coefficient
proportional to $R^{a\ol{b}} p_a p_{\ol{b}}$, where $R_{a\ol{b}}$ is
the Ricci curvature of $M$ \cite{LMT}. Therefore, if $M$ is not
Calabi-Yau, the sigma model in the finite volume is not conformally
invariant. However, in the infinite volume limit the beta-function
vanishes and the theory becomes conformally invariant, even for
manifolds that are not Calabi-Yau.\qed

\end{remark}

\subsection{Correlation functions}    \label{cor fn}

Correlation functions in our model are defined for any Riemann surface
$\Sigma$ with marked points $x_1,\ldots,x_n$, and a collection of
local operators inserted at those points. In a general conformal field
theory with central charge $c=0$ correlation functions are functions
on the moduli space ${\mathcal M}_{g,n}$ of pointed curves
$(\Sigma,(x_i))$.\footnote{we may also need to choose non-zero tangent
vectors, or even germs of local coordinates, at the marked points, but
in the discussion below we will omit them} But our theory carries a
supersymmetry charge $Q$ such that the stress tensor $T(z)$ is
$Q$--exact: $T(z) = [Q,G(z)]_+$, and similarly for the anti-chiral
fields, and so it has the structure of {\em topological conformal
field theory}. In a topological conformal field theory we can
construct not only functions, but also {\em differential forms} on the
moduli space ${\mathcal M}_{g,n}$, by inserting integrals of the
fields $G(z)$ and $\ol{G}(\ol{z})$ (see \cite{W:cs,Zwiebach}). Let us
recall this construction.

Suppose for simplicity that $n>0$, and let ${\mc O}_1,\ldots,{\mc
O}_n$ be some local operators inserted at the points
$x_1,\ldots,x_n$. We will explain how to construct holomorphic
differential forms. The construction is easily generalized to
arbitrary forms. We note that the holomorphic tangent space to the
moduli space ${\mathcal M}_{g,n}$ at $(\Sigma,(x_i))$ is isomorphic to
the double quotient
$$
\Gamma(\Sigma \bs \{ x_1,\ldots,x_n \},T^{1,0} \Sigma) \bs
\bigoplus_{i=1}^n \C(\!(t_i)\!)  \pa_{t_i}/\bigoplus_{i=1}^n \C[[t_i]]
\pa_{t_i},
$$
where $T^{1,0} \Sigma$ is the holomorphic tangent bundle of $\Sigma$
(see, e.g., \cite{FB}, Sect. 17.3, and references therein). Now any
holomorphic vector field on the punctured disc near $x_i$, $\xi_i =
f_i(t_i) \pa_{t_i} \in \C(\!(t_i)\!) \pa_{t_i}$ defines a tangent
vector in $T^{1,0}_{(\Sigma,(x_i))} {\mathcal M}_{g,n}$. To define a
differential $(k,0)$--form on ${\mathcal M}_{g,n}$ corresponding to
${\mc O}_1,\ldots,{\mc O}_n$ we need to describe its values on
$k$--tuples of holomorphic tangent vectors of the above form. Let us
suppose that we have tangent vectors corresponding to the vector fields
$\xi^{(1)}_j,\ldots,\xi^{(\al_j)}_j$ at the point $x_j$. Then, by
definition, the value of this $(k,0)$--form on these tangent vectors
is just the correlation function
$$
\left\langle \prod_{j=1}^n \int \xi^{(1)}_j G(z^{(1)}_j)
\ldots \int  \xi^{(\al_j)}_jG(z^{(\al_j)}_j) {\mc O}_j \right\rangle.
$$
In other words, we ``dress'' the local operator inserted at $x_j$ by
contour integrals of $G(z)$ coupled to the vector fields
$\xi^{(1)}_j,\ldots,\xi^{(\al_j)}_j$. To obtain more general
differential forms, we should use the anti-chiral field
$\ol{G}(\ol{z})$ as well.

If the observables ${\mc O}_j$ have definite fermionic charges, then
among all of these differential forms there is at most one that
is non-zero. Its degree is determined by the corresponding fermionic
charge conservation law.

What do these differential forms look like? Typical observables of the
theory are differential forms on $M$, and $Q$ acts on them as the de
Rham differential. Let ${\mathcal M}_{g,n}(M,\beta)$ be the moduli
space of $(\Sigma,(x_i),\Phi)$, where $\Sigma$ and $(x_i)$ are as
above and $\Phi$ is a holomorphic map $\Sigma \to M$ of degree
$\beta$. Then we have a forgetful map ${\mathcal M}_{g,n}(M,\beta) \to
{\mathcal M}_{g,n}$. Suppose we want to compute the correlation
functions of the local operators corresponding to differential forms
$\omega_i, i=1,\ldots,n$ on $M$, not necessarily closed. Then we
should take the cup product of the pull-backs of the $\omega_i$'s to
${\mathcal M}_{g,n}(M,\beta)$ under the evaluation maps, and take the
push-forward of the resulting differential form to ${\mathcal
M}_{g,n}$. If the $\omega_i$'s are smooth and have compact support,
then one can show that the result is a differential form (not
necessarily of top degree) on ${\mathcal M}_{g,n}$. This is an example
of a correlation function in our conformal field theory. But this is
not the most general example. Other correlation functions correspond
to other local observables, such as the vector fields on $M$ realized
as Lie derivatives acting on differential forms.

Part of this structure is captured by the Gromov-Witten
invariants. Since these moduli spaces ${\mathcal M}_{g,n}(M,\beta)$
are non-compact, we find that if we wish the correlation functions of
$Q$--closed observables (such as closed differential forms on $M$) to
depend only on their cohomology classes, we need to compactify these
moduli spaces. The factorization property of the correlation functions
will then also require that we introduce certain additional components
into the compactified moduli spaces. The Kontsevich moduli spaces
$\ol{\mathcal M}_{g,n}(M,\beta)$ of stable maps provide one with
compactifications which satisfy all desirable properties and are
equipped with the evaluation maps to the target manifold $M$ which one
can use to pull-back differential forms on $M$. \footnote{Note that it
may happen that ${\mathcal M}_{g,n}(M,\beta)$ is empty, but
$\ol{\mathcal M}_{g,n}(M,\beta)$ is non-empty; see the discussion at
the end of \secref{deformation}.} One also has a forgetful map from
$\ol{\mathcal M}_{g,n}(M,\beta)$ to the Deligne-Mumford
compactification $\ol{\mathcal M}_{g,n}$ of ${\mathcal
M}_{g,n}$. Taking the cup product of the pull-backs of such forms
$\omega_i$'s to $\ol{\mathcal M}_{g,n}(M,\beta)$, and then the
push-forward to $\ol{\mathcal M}_{g,n}$, we obtain differential forms
on $\ol{\mathcal M}_{g,n}$ whose cohomology classes now depend only on
the cohomology classes of the $\omega_i$'s. Pairing them with some
natural cohomology classes on $\ol{\mathcal M}_{g,n}$, we obtain the
Gromov-Witten invariants. But since they come from very special
observables of our theory, they correspond to a particular sector of
the full conformal field theory associated to the twisted sigma model
in the infinite volume.

A natural question is how one can see the compactification
$\ol{\mathcal M}_{g,n}(M,\beta)$ of ${\mathcal M}_{g,n}(M,\beta)$ in
the framework of the conformal field theory with the action
\eqref{action infty}. A possible answer is that the integrals over the
additional strata may naturally appear when one performs a
regularization of the integral over the zero modes of the $p_a$'s and
$p_{\ol{a}}$'s along the lines described above.

Another part of this structure has been studied in mathematical
literature starting with \cite{MSV}. It is encoded by a sheaf of
chiral algebras over $M$, called the chiral de Rham complex, which is
defined by gluing the free chiral algebras on the overlaps of the open
subsets. From the point of view of the sigma model, this chiral
algebra corresponds to the cohomology of the right moving supercharge
of the twisted sigma model in the {\em perturbative} regime (i.e.,
without counting instanton contributions), as explained in
\cite{W:new,Kapustin}. However, the knowledge of this cohomology is
not sufficient for determining the correlation functions of the sigma
model. In order to determine them one needs to generalize the
construction of this chiral algebra to the full conformal field theory
and to include the instanton corrections. This is done in this paper
in the case when the target manifold is a toric variety.

The idea is to realize the quantum field theory governed by the action
\eqref{action infty} in the case when the target manifold $M$ is a
toric variety as a {\em deformation of a free field theory}. A toric
variety ${\mathbb P}_S$ has a particularly nice open cover $\{
{\mathbb A}_{\sigma(i)} \}_{i=1,\ldots,N}$ with each open subset
${\mathbb A}_{\sigma(i)}$ isomorphic to $\C^d$ and their intersection
$\TT_S$ to $(\C^\times)^d$ (see \secref{recollections}). The
complement of $\TT_S$ in ${\mathbb P}_S$ is a divisor with components
$C_i$ equal to the complements of ${\mathbb A}_{\sigma(i)}$ in
${\mathbb P}_S$. Our idea is that the sigma model corresponding to a
target manifold $M$ is equivalent to a deformation of the sigma model
with the target manifold $M \bs C$, where $C$ is a divisor, by means
of a marginal vertex operator determined by $C$. Now, starting with
the sigma model with the target $\TT_S$, which is a free field theory,
we may build the sigma models with the target manifolds obtained by
gradually ``gluing'' back the divisors $C_i$. Each time we ``glue''
back a divisor $C_i$, we deform the theory by a vertex operator
corresponding to $C_i$. Thus, the end result, which is the sigma model
with the target ${\mathbb P}_S$, is identified with the deformation of
the free field theory associated to $\TT_S$ by means of the vertex
operators corresponding to all $C_i, i=1,\ldots,N$. In this paper we
identify these vertex operators and construct these deformations
explicitly. Moreover, we use this description of the sigma model of
${\mathbb P}_S$ to give a new interpetation of mirror symmetry.

We expect that one can give a similar description to the sigma models
corresponding to more general target manifolds. A general complex
manifold can be covered by open subsets that are analytically
isomorphic to domains in $\C^n$. The supersymmetric sigma model
corresponding to each of this open subsets is described by a free
field theory which may be viewed as a system of decoupled bosonic and
fermionic ghosts. So one may hope to define the quantum theory for a
general K\"ahler target manifold $M$ by appropriately ``gluing''
together the free field theories corresponding to these open
subsets. The mathematical works on the chiral de Rham complex indicate
that this is a non-trivial task which requires methods that up to now
have not been widely used by physicists in this context, such as Cech
cohomology. However, for toric varieties our task is considerably
simplified by the existence of a particularly nice cover. We will use
this cover in order to realize the sigma model as a deformation of a
free field theory.

To illustrate these ideas, we will now consider the case when the
target manifold $M$ is $\pone$.

\subsection{Warm-up example: From $\C$ to $\pone$}    \label{warmup}

As a warm-up example, we will consider the case of the target manifold
$M=\pone$. The corresponding non-linear sigma model will be defined as
a deformation of the linear model with the target $\C$. In the next
section we will define the same non-linear model as a deformation of
the linear model with the target $\C^\times$, which we will find to be
technically more convenient. However, it is instructive to start by
looking first at the deformation from $\C$ to $\pone$.

The theory with the target $\C$ is a free conformal field
theory with the chiral fields $X(z), p(z)$, $\psi(z)$, $\pi(z)$
and their anti-chiral partners with the action \eqref{action first
time}. The chiral fields obey the standard OPEs
$$
p(z) X(w) = - \frac{i}{z-w} + \on{reg.}, \qquad
\psi(z) \pi(w) = - \frac{i}{z-w} + \on{reg.}
$$
This is nothing but the free theory of bosonic and fermionic ghosts
(also known as a $\beta\gamma$--system and a $bc$--system), and its
quantization is relatively straightforward.

We wish to interpret holomorphic maps $\Sigma \to \pone$ within the
framework of this free field theory. Namely, we view such maps as
meromorphic maps $\Sigma \to \C$. Let $w_1,\ldots,w_n$ be the points
of $\Sigma$ where this map has a pole. Generically, all these poles
will be of order one. As explained in the introduction, our proposal
is that we can include such maps by inserting in the correlation
functions of the linear sigma model certain vertex operators at the
points $w_1,\ldots,w_n$. In the case at hand, we propose the following
candidate for this operator:
$$D(z,\ol{z}) = \delta^2(p)(z,\ol{z}) \pi(z) \ol{\pi}(\ol{z}).$$

What is the meaning of the operator $\delta^2(p)(z,\ol{z})$ from the
Lagrangian point of view? Recall that the field $p(z)$ is a Lagrange
multiplier responsible for the equation of holomorphy $\pa_{\ol{z}} X
= 0$. Therefore the insertion of the field $\delta^2(p)(z,\ol{z})$ in
the path integral is the instruction to relax this equation at the
point $z \in \Sigma$ in the minimal possible way. This just means that
our map $X: \Sigma \to \pone$ should cease to be holomorphic at the
point $z \in \Sigma$, i.e., it should develop a pole. We need to
multiply $\delta^2(p)(z,\ol{z})$ by its odd counterpart, namely
$\delta^2(\pi)(z,\ol{z})$, which is nothing but the operator $\pi(z)
\ol{\pi}(\ol{z})$. This gives us the above operator $D(z,\ol{z})$.
Inserting the operators $D(w_i,\ol{w}_i), i=1,\ldots,n$, corresponds
to considering meromorphic maps $\Sigma \to \C$ with poles precisely
at the points $w_1,\ldots,w_n$, or equivalently, considering the
holomorphic maps $\Sigma \to \pone$ which pass through the point
$\infty \in \pone$ precisely at the points $w_1,\ldots,w_n \in
\Sigma$.

The operator $D(z,\ol{z})$ also has a transparent meaning from the
point of view of the operator formalism. While operators of the form
$\delta^2(X)(z,\ol{z})$ are quite common, the operators
$\delta^2(p)(z,\ol{z})$ may appear at first glance as somewhat more
exotic. But the mystery disappears if one considers the corresponding
state in the Hilbert space of the linear sigma model corresponding to
a small circle around a point $z \in \Sigma$. To simplify notation,
set $z=0$. Then this space contains the direct sum of the the tensor
products
$$
F_N \otimes \ol{F}_{N}, \qquad N \in \Z,
$$
of the Fock representations $F_N$ the Heisenberg algebra generated by
the Fourier modes of the chiral fields $$X(z) = \sum_{n \in \Z} X_n
z^{-n}, \qquad p(z) = \sum_{n \in \Z} p_n z^{-n-1},$$ and their
anti-holomorphic analogues $\ol{F}_{N}$. The vacuum vector $\vac
\otimes \ol{\vac}$ is in $F_0 \otimes \ol{F}_{0}$. The vector
$\vac \in F_0$ is annihilated by $\int X(z) f(z) dz$ for all
holomorphic one-forms $f(z) dz$ on a small disc around $0$, where the
integral is taken over a small circle around $0$ (i.e., it is
annihilated by $X_n, n>0$) and by $\int p(z) g(z) dz$, for all
holomorphic functions $g(z)$ on the small disc around $0$ (i.e., it is
annihilated by $p_n, n \geq 0$). The vector $\ol{\vac}$ satisfies
similar equations.

Now, the vector corresponding to the operator $\delta^2(p)(0,0)$ is
nothing but the tensor product of the highest weight vectors from
other Fock spaces, namely, $|1\rangle \otimes \ol{|1 \rangle} \in F_1
\otimes \ol{F}_{1}$. The vector $|1 \rangle$ satisfies
$$
\int X(z) f(z) dz \cdot |1 \rangle = 0, \qquad f(z) \in z \C[[z]],
$$
$$
\int p(z) g(z) dz \cdot |1 \rangle = 0, \qquad g(z) \in z^{-1}
\C[[z]].
$$
In other words, $|1 \rangle$ is annihilated by $X_n, n>1$, and by
$p_n, n\geq -1$. So $$\delta^2(p)(z,\ol{z}) = \delta(p)(z)
\delta(\ol{p})(\ol{z}),$$ where $\delta(p)(z)$ is nothing but the
chiral field corresponding to the highest weight vector $|1\rangle$ of
the Fock representation $F_1$ of the Heisenberg algebra, and
$\delta(\ol{p})(\ol{z})$ is its anti-chiral analogue corresponding to
the anti-chiral state $\ol{|1 \rangle}$.

Likewise, $\pi_{-1}\ol{\pi}_{-1}\vac$ is a highest weight vector over
the Clifford algebra generated by the Fourier coefficients of the
fields $\psi(z), \pi(z), \ol{\psi}(z), \ol{\pi}(z)$. It is annihilated
by $\psi_n, \ol\psi_n, n>1$, and $\pi_n, \ol\pi_n, n \geq -1$.

Incidentally, from this point of view $\delta^2(X)(z,\ol{z})$ is nothing
but the operator corresponding to the state $|-1 \rangle \otimes
\ol{|-1 \rangle}$. So the familiar operator $$\OO_0(z,\ol{z}) =
\delta^2(X)(z,\ol{z}) \psi(z) \ol{\psi}(\ol{z})$$ is an analogue of our
operator $D(z,\ol{z})$, which may in fact be used to represent the
observable in the Gromov-Witten theory corresponding to the degree two
cohomology class of $\pone$.

The conformal dimension of the field $\delta^2(p)$ is $(-1,-1)$.  This
is in fact a special case of a general fact: if $\Phi(z,\ol{z})$ is a
bosonic field of conformal dimension $(\Delta,\ol\Delta)$ and charge
$\nu$, then $\delta^2(\Phi)(z,\ol{z})$ should have conformal dimension
$(-\Delta,-\ol\Delta)$ and charge $-\nu$. Note also that $D(z,\ol{z})$
has conformal dimension $(0,0)$.

Let us compute the correlation function of these observables for
$\Sigma$ of genus zero. From the Gromov-Witten theory we know that the
correlation function is non-zero if the number of insertions is odd,
$2n+1$, and then the answer should be equal to $q^n$, because it
corresponds to holomorphic maps of degree $n$. Let us explain how to
reproduce exactly this answer within the framework of the linear sigma
model. Observe that a map of degree $n$ has to pass through $\infty$
exactly $n$ times (with multiplicities, in general, but generically
the multiplicities will all be equal to one). This means that we have
to insert the operator $D(z,\ol{z})$ at $n$ distinct points. But this
operator has charge $1$ (with respect to the current $\Wick X(z)p(z)
\Wick$) and ghost number $1$, while the operator $\OO(z,\ol{z})$ has
charge $-1$ and ghost number $-1$. The anomalous conservation law in
genus zero demands that the total charge and the ghost number be both
equal to $-1$. Therefore in order to compensate for the $n$ insertions
of the operator $D(z,\ol{z})$ we have to insert the operator
$\OO(z,\ol{z})$ at $n+1$ additional points. After that we reproduce
the answer of the Gromov-Witten theory because the correlation
function of these operators is equal to $1$, which we should multiply
by $q^n$ to account for the degree of the map. In other words, in
order to account for the degrees of the holomorphic maps we should
really be inserting the operator $q D(z,\ol{z})$ rather than
$D(z,\ol{z})$.

In the Gromov-Witten theory one also considers the fields obtained by
cohomological descent from the basic fields described above (see
\cite{W:mirror}). The cohomological descendants of an operator
$\OO$ satisfy the equations
$$
d\OO = [Q_{\on{tot}},\OO^{(1)}], \qquad d\OO^{(1)} =
[Q_{\on{tot}},\OO^{(2)}],
$$
where $Q_{\on{tot}} = Q + \ol{Q}$ is the supersymmetry charge. To
calculate them, we observe that we have two (twisted) $N=2$
superconformal algebras with the chiral one generated by the fields
$$
G(z) = i \pa_z X(z) \pi(z), \qquad Q(z) = - i p(z) \psi(z),
$$
$$
T(z) = - i \Wick \pa_z X(z) p(z) \Wick - i \Wick \pi(z) \pa_z \psi(z)
\Wick, \qquad J(z) = i \Wick \psi(z) \pi(z) \Wick \, ,
$$
and similarly for the anti-chiral one. The chiral supersymmetry charge
is the operator $Q = \int Q(z) dz$, and $G(z)$ satisfies $$\int Q(w)
dw \cdot G(z) = T(z)$$ (here and below, in similar formulas, the
contour of integration goes around $z$, and we suppress the factor
$1/2\pi i$). In particular, we have
$$[Q,G_{-1}]_+ = L_{-1},$$ where $G_{-1} = \int G(z) dz$. We have
similar formulas for $\ol{Q}$.

This allows us to find $\OO^{(1)}$ and $\OO^{(2)}$ from the formulas
$$
\OO^{(1)} = G_{-1} \cdot \OO dz + \ol{G}_{-1} \cdot \OO d\ol{z},
\qquad \OO^{(2)} = G_{-1} \ol{G}_{-1} \cdot \OO dz d\ol{z},
$$
provided that $Q_{\on{tot}} \cdot \OO = 0$. In particular, since
$Q_{\on{tot}} \cdot D(z,\ol{z}) = 0$, we find that
$$
D^{(2)}(z,\ol{z}) = \left( \int X(w) w dw \int \ol{X}(\ol{w}) \ol{w}
d\ol{w} \cdot \delta^2(p)(z,\ol{z}) \right) \pi(z) \pa_z \pi(z)
\ol\pi(\ol{z}) \pa_{\ol{z}} \ol\pi(\ol{z}) dz d\ol{z}.
$$
The bosonic part of this field corresponds to the state $X_{1}
|1\rangle \otimes\ol{X}_1 \ol{|1\rangle} \in F_1 \otimes
\ol{F}_1$. Note that the field $D^{(2)}(z,\ol{z})$ has conformal
dimension $(1,1)$.

In the setting of the linear sigma model the maps $\Sigma \to \pone$
of degree $n$ are the same as meromorphic maps with poles at $n$
points (counted with multiplicity). As we argued above, those should
be counted via the insertion of the vertex operator
$qD(z,\ol{z})$. Since we will be integrating over all such maps, and
hence over all possible positions of the poles, the degree $n$
contribution to the correlation function $\langle {\mc O}_1 \ldots
{\mc O}_n \rangle_{\pone}$ of local observables in the non-linear
sigma model with the target $\pone$ (such as $\OO_0(z,\ol{z})$
introduced above) should be equal to the correlation function of these
operators in the linear model with the additional insertion of the
integral of the $(1,1)$--forms $q D^{(2)}(z,\ol{z})$, obtained by
cohomological descent from the operators $D(z,\ol{z})$ introduced
above. Thus, this correlation function should be given by
$$
\sum_{n=0}^\infty \frac{q^n}{n!} \langle {\mc
  O}_1 \ldots {\mc O}_m \int
D^{(2)}(w_1,\ol{w}_1) \ldots \int D^{(2)}(w_1,\ol{w}_n) \rangle_{\C}
$$
(the $1/n!$ factor is due to the fact that the points $w_1,\ldots,w_n$
are unordered). But this is the same as the correlation function in
the linear sigma model deformed by the marginal operator $q
D^{(2)}(z,\ol{z})$.

This suggests that {\em the non-linear sigma model with the target
$\pone$ in the infinite volume limit is equivalent to the linear sigma
model deformed by the marginal operator $D^{(2)}(z,\ol{z})$}, i.e.,
the theory defined by the action
$$
\frac{1}{2\pi} \int_{\Sigma} d^2 z \; \left( i p \pa_{\ol{z}} X + i
\pi \pa_{\ol{z}} \psi + i \ol{p} \pa_z \ol{X} + i \ol\pi \pa_z \ol\psi
+ q D^{(2)} \right).
$$

While nice and intuitive, the representation of the deforming operator
in terms of the delta-function $\delta^2(p)(z,\ol{z})$ is rather
inconvenient for practical calculations. One possible way to do that
is to invoke the Friedan-Martinec-Shenker bosonization \cite{FMS} of
the $p,X$ system:
$$
X(z) = e^{u(z)+v(z)}, \qquad p = - \pa v(z) e^{-u(z)-v(z)},
$$
where $u(z)$ and $v(z)$ are the scalar fields having the OPEs
$$
u(z) u(w) \sim - \log(z-w), \qquad v(z) v(w) \sim \log(z-w).
$$
We have similar formulas for the anti-chiral fields $\ol{p},\ol{X}$.
Then we have the following bosonic representation $$\delta(p)(z) =
e^{u(z)}, \qquad \delta(\ol{p})(\ol{z}) = e^{\ol{u}(\ol{z})}.$$ It is
easy to see that these fields have the right OPE with the fields
$X(z), p(z)$ and their complex conjugates. Since the conformal
dimension of $e^{\al u(z)}$ is $-\al(\al+1)/2$, we obtain that the
conformal dimension of $\delta(p(z))$ is indeed $-1$.

Thus, we obtain the following realization of the fields introduced above:
\begin{align*}
D(z,\ol{z}) &= e^{u(z)+\ol{u}(\ol{z})} \pi(z) \ol\pi(\ol{z}), \\
D^{(2)}(z,\ol{z}) &= e^{2(u+\ol{u})+(v+\ol{v})} \pi \pa_z \pi \ol{\pi}
\pa_{\ol{z}} \ol{\pi} dz d\ol{z}.
\end{align*}
However, the FMS bosonization identifies the $X,p$ system with a
subalgebra of the chiral algebra of the two scalar bosons $u,v$. To
get an isomorphism, we need to invert $X$, i.e., pass from $\C$ to
$\C^\times$ (see \cite{FF:weil}). This already indicates that it is
more convenient to formulate the theory on $\C^\times$ rather than on
$\C$. This leads us to the toric sigma model introduced in the next
section.

\section{The model with the target $\C^\times$}    \label{toric sigma
  model}

\subsection{Toric sigma model}    \label{toric sigma}

We would like to express the correlation functions of the sigma model
with the target $\pone$ in the limit of infinite volume in terms of
the operator formalism of the sigma model with the target $\C^\times =
\pone \bs \{ 0,\infty \}$, also at the infinite volume.

To define the sigma model with the target $\C^\times$ we will
use the logarithmic coordinate $X = R + i \phi$, where $\phi$ is
periodic with the period $2\pi$. In other words, we identify
$\C^\times$ with $\R \times S^1$, where $R$ is a coordinate on $\R$
and $\phi$ is a coordinate on $S^1$. We introduce the metric
\begin{equation}    \label{cyl metric}
t(dR^2 + d\phi^2) = t dX d\ol{X},
\end{equation}
so the circle has radius $\sqrt{t}$. The action of the sigma model in
the first order formalism, introduced in \secref{first order}, is
\begin{equation}    \label{IR}
I_t = \frac{1}{2\pi} \int_{\Sigma} d^2 z \; \left( i p \pa_{\ol{z}} X +
i \ol{p} \pa_z \ol{X} + i \pi \pa_{\ol{z}} \psi + i \ol\pi \pa_z
\ol\psi + t^{-1} p \ol{p} \right).
\end{equation}
To eliminate $p$ and $\ol{p}$ in the path integral by completing the
action to a square and integrating them out (see \remref{misleading}),
we substitute the following expressions in the Lagrangian:
\begin{equation}    \label{eqs of motion}
p = - i t \pa_z \ol{X}, \qquad \ol{p} = - i t \pa_{\ol{z}} X.
\end{equation}
Then we obtain the usual action of the sigma model with the target $\R
\times S^1$ with the metric \eqref{cyl metric}:
$$
\frac{t}{2\pi} \int_{\Sigma} d^2 z \; \left( \pa_{\ol{z}} X \pa_z
\ol{X} + i \pi \pa_{\ol{z}} \psi  + i \ol\pi \pa_z \ol\psi \right).
$$

In the limit $t \to \infty$ the last term in $I_t$ drops out and we
obtain the action
\begin{equation}    \label{action infinity}
\frac{i}{2\pi} \int_{\Sigma} d^2 z \; \left( p \pa_{\ol{z}} X + \pi
\pa_{\ol{z}} \psi + \ol{p} \pa_z \ol{X} + \ol\pi \pa_z \ol\psi
\right).
\end{equation}
We call this model a {\em toric sigma model} with the target
$\C^\times$.

Equations of motion imply that fields $X(z), p(z), \psi(z), \pi(z)$
are holomorphic ($X(z)$ and $\psi(z)$ have conformal dimension $0$ and
$p(z), \pi(z)$ have conformal dimension $1$), while their complex
conjugates $\ol{X}(\ol{z}), \ol{p}(\ol{z}), \ol\psi(\ol{z}),
\ol\pi(\ol{z})$ are anti-holomorphic. They obey the standard OPEs
\begin{equation}    \label{pX OPE}
X(z) p(w) = - \frac{i}{z-w} + \Wick X(z) p(w) \Wick, \qquad
\psi(z) \pi(w) = - \frac{i}{z-w} + \Wick \psi(z) \pi(w) \Wick,
\end{equation}
and similarly for the anti-chiral fields. So this is a free field
theory which is the toric version of the well-known system of bosonic
and fermionic ghost fields. It possesses an $N=2$ superconformal
symmetry. The generating fields of the left moving $N=2$ (twisted)
superconformal algebra are given by the following formulas:
\begin{equation}    \label{N=2}
Q(z) = - i p(z) \psi(z) - \pa_z \psi(z), \qquad G(z) = i \pa_z X(z)
\pi(z),
\end{equation}
$$
T(z) = - i \Wick \pa_z X(z) p(z) \Wick - i \Wick \pi(z) \pa_z \psi(z)
\Wick, \qquad J(z) = i \Wick \psi(z) \pi(z) \Wick \, + \pa_z X(z).
$$
There are also anti-chiral fields $\ol{Q}(\ol{z}), \ol{G}(\ol{z}),
\ol{T}(\ol{z})$, and $\ol{J}(\ol{z})$, given by similar formulas,
which generate the right moving copy of the $N=2$ superconformal
algebra.

The Hilbert space of the theory is built from bosonic Fock
representations of the Heisenberg algebra generated by the Fourier
coefficients of the fields $\pa_z X(z), p(z)$, $\pa_{\ol{z}} X(\ol{z}),
\ol{p}(\ol{z})$ and fermionic Fock representations of the Clifford
algebra generated by the Fourier coefficients of $\psi(z), \pi(z),
\ol\psi(\ol{z}), \ol\pi(\ol{z})$.  The precise structure of the
bosonic Hilbert space and the state-field correspondence will be
described in \secref{hamiltonian}. Here we focus on the most salient
features of the theory.

\subsection{Holomorphic vortices}    \label{holomortex}

In the canonical quantization of the toric sigma model we consider the
theory defined on the cylinder $\Sigma$, with the holomorphic
coordinate $z = e^{t+is}, t \in \R, s \in \R/2\pi\Z$. Because our
target space is also a cylinder, we find that we can allow non-trivial
winding, i.e., we can allow $X(e^{2\pi i}z)$ to differ from $X(z)$ by
a an integral multiple of $2\pi i$. This, together with the condition
of holomorphy, means that $X(z)$ and $\ol{X}(\ol{z})$ may be written
as follows:
$$
X(z) = \om \log z + \sum_{n \in \Z} X_n z^{-n}, \qquad \ol{X}(\ol{z}) =
\om \log \ol{z} + \sum_{n \in \Z} \ol{X}_n \ol{z}^{-n},
$$
where $\om$ is the {\em winding operator} which is allowed to take
integer values. This indicates that the Hilbert space may contain
states that have non-zero value of the operator $\om$, and hence
non-zero winding.

A convenient way to understand this is by interpreting the toric sigma
model as the $\Z$--orbifold of the corresponding model with the target
$\C$. The latter is the free field theory that we discussed in
\secref{warmup}. It is described by the action \eqref{action
infinity}, where how $X(z)$ and $\ol{X}(\ol{z})$ are
single-valued. The group $\Z$ is a symmetry group of the action,
shifting $X$ by integer multiples of $2\pi i$. We expect that our
toric sigma model with the target $\C/2\pi i \Z$ may be obtained from
the corresponding theory with the target $\C$ by taking its
$\Z$--orbifold. The corresponding twist fields should then be exactly
the fields with non-zero winding number $\omega$.

This is analogous to the fact that the vortex operators of the sigma
model with the target $\C/2\pi i \Z$ at the finite radius may be
interpreted as the twist fields arising in the $\Z$--orbifolding of
the usual linear sigma model with the target $\C$. Because of this
analogy, we call the twist fields arising in the toric sigma model
{\em holomortex operators}. However, the vortex operators and the
twist fields that we have at the infinite radius have different
nature.

To explain this point, it is convenient to work in the logarithmic
coordinates $s,t$ on the worldsheet cylinder $\Sigma_0$. In the finite
volume theory the coordinates $R,\phi$ on the target cylinder $M_0$
are completely independent, and the winding occurs in the $\phi$
variable, independently of $R$. In other words, there are harmonic
maps $X: \Sigma_0 \to M_0$ which are constant along $R$, but wind
around $\phi$, such as $R = 0, \phi = m s$, where $m \in \Z$. The
vortex operator with the winding number $m$ belongs to the sector of
the theory corresponding to maps of this type. But in the infinite
volume limit the map $X: \Sigma \to M$ has to be {\em
holomorphic}. Therefore $R$ and $\phi$ are no longer independent. Now
we have maps of the form $R = m t, \phi = m s$, where $m \in \Z$, so
$R$ as well as $\phi$ depend on $(s,t)$. That is why there is no
straightforward way to define the holomorphic winding operators as a
naive limit of the vortex operators in the infinite volume limit.

What are then the explicit formulas for the holomortex operators?
Denoting the operator with the winding number $m$ by
$\Psi_m(z,\ol{z})$, we find that we need to have the following OPEs:
\begin{align}    \label{Psi}
X(z) \Psi_m(w,\ol{w}) &= m\log(z-w) \Psi_m(w,\ol{w}) + ..., \\ \notag
\ol{X}(\ol{z}) \Psi_m(w,\ol{w}) &= m\log(\ol{z}-\ol{w})
\Psi_m(w,\ol{w}) + ...
\end{align}

Using the OPEs \eqref{pX OPE}, we find that the field
\begin{equation}    \label{Phim1}
\Psi_m(z,\ol{z}) = e^{- i m\int P}(z,\ol{z}) = \exp \left( - i m
\int_{z_0}^z (p(w) dw + \ol{p}(\ol{w}) d\ol{w}) \right)
\end{equation}
(or any of its scalar multiples) has precisely the OPEs \eqref{Psi}
with $X(z)$ and $\ol{X}(\ol{z})$.

Formula \eqref{Phim1} a priori depends on the point $z_0$ and the
integration contour. We will give a more precise definition of these
operators acting on the Hilbert space of the theory in
\secref{hamiltonian}. Here we would like to comment that for the
purposes of this paper we only need to consider the correlation
functions of the operators $e^{\pm i \int P}$. We will postulate that
a correlation function of such operators will be non-zero if and only
if an equal number of these operators with the $+$ and $-$ signs are
involved. (In \secref{T-duality} we will see that this condition
naturally comes from integrating over the zero mode of the dual
variable $U$.) Then we simply define the correlation function by
pairing the $+$ and $-$ operators in an arbitrary way and integrating
over the contours going from the location of the $-$ operator to the
location of the $+$ operator in each pair. The result is independent
of the choice of the pairing as long as all other operators in the
correlation function have well-defined OPEs with the operators $e^{\pm
i \int P}$, as discussed below. Note also that while the individual
operator $e^{\pm i \int P}$ is a priori defined only up to a scalar
multiple, once we normalize one of them, the other is also
automatically normalized. Therefore the product of an equal number of
the $+$ and $-$ holomortex operators does not depend on the choice of
normalization. This gives us a well-defined prescription for the
computation of the correlation functions that we need, and it is easy
to generalize it to the correlation functions involving the fields
$\Psi_m$ with $m \neq \pm 1$.

The presence of the holomortex operators $e^{i m \int P}, m \in \Z$,
given by formula \eqref{Phim1}, in our theory places restrictions on
what other fields are allowed. Namely, those fields must have
well-defined OPEs with the fields $e^{i m\int P}, m \in \Z$. (This
insures the contour independence of the correlation functions
discussed in the previous paragraph.) This is analogous to the case of
the sigma model with the target $\C/2\pi i \Z$ at the finite radius,
considered as a $\Z$--orbifold. In the linear sigma model we have the
fields $e^{ir\phi}$ with arbitrary $r \in \R$, but after orbifolding
$r$ is quantized and can take only integer values. This condition
insures that these fields have well-defined OPEs with the vortex
operators, which are the orbifold twist fields at the finite radius.

Let us analyze what conditions are imposed by the presence of the
twist fields $e^{im\int P}, m \in \Z$ in our theory. The fields $p(z),
\ol{p}(\ol{z})$ have well-defined OPEs with them, and so do the
derivatives $\pa_z X(z), \pa_{\ol{z}} \ol{X}(\ol{z})$. Next, we look
at the exponential fields $\exp(\al X(z) + \beta
\ol{X}(\ol{z}))$. They have the following OPEs with $e^{- i m\int
P}(w,\ol{w})$:
\begin{multline*}
\exp(\al X(z) + \beta \ol{X}(\ol{z})) e^{- i m\int P}(w,\ol{w}) = \\
(z-w)^{m\al} (\ol{z}-\ol{w})^{m\beta} \Wick \exp(\al X(z) + \beta
\ol{X}(\ol{z})) e^{- i m\int P}(w,\ol{w}) \Wick \, .
\end{multline*}
The condition for the right hand side to be single-valued is that $\al
- \beta \in \Z$. This condition ensures that the correlation functions
of the allowed operators and the operators $e^{- i m\int P}(w,\ol{w})$
do not depend on the choice of the contours of integration. The
operator content of the theory is described in more detail in
\secref{hamiltonian}.

\subsection{$T$--duality}    \label{T-duality}

Now we will show that the toric sigma model introduced in the previous
section is equivalent to the ordinary sigma model with the target space
being the torus $\R \times S^1$ equipped with the {\em Minkowski}
metric such that the circle is {\em isotropic}. In this realization
the holomortex operators $e^{i m \int P}$ have a particularly simple
form.

In this section we discuss the path integral realization of the
duality. The operator realization will be considered in
\secref{T-dual, operator}. Let us introduce the one-form
$$
P = p(z) dz + \ol{p}(\ol{z}) d\ol{z}
$$
on $\Sigma$. We choose the real structure in which the complex
conjugate of $p$ is $\ol{p}$, so that the one-form $P$ is real. Then
we rewrite the bosonic part of the action \eqref{action infinity} as
follows:
\begin{equation}    \label{bosonic part}
I_{\on{bos}} = \frac{i}{2\pi} \int_\Sigma (- P \wedge d \phi + P
\wedge * dR).
\end{equation}
Here $*$ denotes the Hodge star operator on $\Sigma$, which in
coordinates looks as follows: $* dz = -idz, *d\ol{z} = i
d\ol{z}$. Recall that our convention for the integration measure on
$\Sigma$ is $d^2 z = i dz \wedge d\ol{z}$.

Let us integrate out the field $\phi$ in the path integral. Then we
obtain the constraint $dP = 0$, or in components $\pa_z \ol{p} =
\pa_{\ol{z}} p$. A general solution of this equation is
\begin{equation}    \label{P}
P = dU = dU_0 + \sum_{j \in I} a_j \omega_j,
\end{equation}
where $U_0$ is a real single-valued field and the $\omega_i$'s are
closed real one-forms representing a basis in the first cohomology
group of $\Sigma$. We choose them in such a way that they are harmonic
and their integrals over cycles in $\Sigma$ are integers and $J^{kl} =
\int \omega_k \wedge \omega_l$ is an integral skew-symmetric matrix
with determinant one. We claim that the coefficients $a_i$ are
constrained to be of the form $a_j = 2\pi m_j, m_j \in \Z$, and so
$U$ is a $2\pi$--periodic field.

We follow the presentation of the book \cite{Mirror}, Sect.~11.2. The
field $\phi$ takes values in $\R/2\pi \Z$ and therefore it is allowed
to have non-trivial winding. This means that $d\phi$ may be expressed
by the formula
$$
d\phi = d\phi_0 + 2\pi \sum_{i \in I} n_i \omega_i, \qquad n_i \in
\Z,
$$
where $\phi_0$ is a real single-valued function. Then we have
$$
\frac{1}{2\pi} \int_\Sigma P \wedge d\phi = \sum_{i,j \in I}
a_i J^{ij} n_j.
$$
Taking the summation over the $n_j$'s in the path integral, we find
from the Poisson summation formula that $a_j = 2\pi m_j, m_j \in
\Z$. Hence $U$ is a function $\Sigma \to \R/2\pi\Z$.

Thus, we have the following transformation
formulas:
$$
p(z) = \pa_z U(z,\ol{z}), \qquad \ol{p}(\ol{z}) = \pa_{\ol{z}}
U(z,\ol{z}),
$$
$$
\frac{1}{2}(X(z) + \ol{X}(\ol{z})) = R(z,\ol{z}).
$$
These formulas are closely related to the Friedan-Martinec-Shenker
bosonization discussed in \secref{warmup}. The holomortex operators
$e^{i m \int P}$ have a particularly simple realization in the dual
variables:
$$
e^{i m \int_{z_0}^z P} = e^{i m U(z)} e^{-i m U(z_0)},
$$
and this is the reason why the dual theory will be convenient for our
purposes.

Let us introduce the improved holomortex operators $e^{i m U(z)}$. The
integration over the zero mode of the field $U(z)$ will guarantee that
the correlation function of the operators $e^{\pm i U(z)}$ will be
non-zero if and only if equal numbers of the operators $e^{i U(z)}$
and $e^{-i U(z)}$ are involved. This is precisely the condition that
we imposed by hand in \secref{holomortex}.\footnote{We thank V. Lysov
for a discussion of this point.} On the other hand, if this condition
is satisfied, then the correlation functions of the improved
holomortex operators are the same as the correlation functions of the
original ones. Hence from now on we will use the improved holomortex
operators in our computations.

The dual theory is formulated in terms of the fields $U$ and $R$
with the action
\begin{equation}    \label{bos action}
\wt{I}_{\on{bos}} = \frac{i}{2\pi} \int_\Sigma dU \wedge * dR =
\frac{i}{2\pi} \int_\Sigma d^2 z \; (\pa_z U \pa_{\ol{z}} R +
\pa_{\ol{z}} U \pa_z R).
\end{equation}
This is the action of the sigma model with the target the cylinder $\R
\times (\R/2\pi \Z)$ with coordinates $(R,U)$, but with the {\em
Minkowski} metric $idR dU$. Note that the compact direction $U$ is
isotropic, and so the notion of the ``radius'' of this cylinder does
not make sense.

Since the one-forms $\omega_j$'s in formula \eqref{P} are chosen to be
harmonic, we can replace in the action the multivalued function $U$ by
the single-valued function $U_0$. However, we then have to remember to
integrate in the path integral not only over $U_0$ but also sum up
over all possible values of $a_j = 2\pi m_j, m_j \in \Z$. Because our
metric is Minkowski and the dual variable to $U$, namely $R$, is
non-periodic, this leads to some non-trivial consequences as discussed
below in \secref{dual}.

The fermionic part of action of the theory remains the same, so the
total action of the dual theory is
\begin{equation}    \label{total action}
\wt{I} = \frac{i}{2\pi} \int_\Sigma d^2 z \; (\pa_z U
\pa_{\ol{z}} R + \pa_{\ol{z}} U \pa_z R + \pi \pa_{\ol{z}} \psi
+ \ol\pi \pa_z \ol\psi).
\end{equation}

Note that when we deform the action \eqref{action infinity} to a
finite radius $r$, we add to it the term $\frac{1}{r^2} p \ol{p}$,
which in the dual variables looks as $\frac{1}{r^2} \pa_z U
\pa_{\ol{z}} U$. Therefore we see that the metric on the torus is
changing in such a way that the circle in the $U$ direction acquires
radius $r^{-1}$, as we should expect under $T$--duality at the finite
radius.

\section{Changing the target from $\C^\times$ to $\pone$}
\label{changing to pone}

The correlation functions of the toric sigma model correspond to path
integrals over all maps $\Sigma \to \C^\times$. Then, since the path
integral over $p$ and $\pi$ and their complex conjugates is
interpreted as the delta-form supported on the holomorphic maps, as we
argued above, any correlation function of the fields involving $X(z)$
and $\psi(z)$ (and their complex conjugates) may be written in terms
of the holomorphic maps $\Sigma \to \C^\times$, which are necessarily
constant for compact $\Sigma$. Therefore the correlation functions
reduce to integrals over the zero mode (i.e., over the image of the
constant map $\Phi: \Sigma \to \C^\times$). Is it possible to
interpret holomorphic maps $\Sigma \to \pone$ within the framework of
the toric sigma model?

\subsection{Deformation of the toric sigma model}
\label{deformation}

As we explained in the Introduction, holomorphic maps $\Sigma \to
\pone$ may be viewed as holomorphic maps $\Sigma \bs \{ w_i^{\pm} \}
\to \C/2\pi i \Z$ with logarithmic singularities at some points
$w^\pm_1,\ldots,w^\pm_N$ where this map behaves as $\pm
\log(z-w^\pm_i)$. These singular points correspond to zeroes and poles
of $\exp \Phi$, and generically they will be distinct. Our proposal is
that {\em we can include these maps by inserting in the correlation
function of the linear sigma model certain vertex operators}
$\Psi_\pm(w^\pm_i)$.

The defining property of the operators $\Psi_\pm(w)$ is that their
operator product expansion (OPE) with $X(z)$ should read
$$
X(z) \Psi_\pm(w) = \pm \log(z-w) \Psi_\pm(w).
$$
We have already found such operators in \secref{holomortex}. These are
the holomortex operators
$$
\Psi_\pm(w) = e^{ \mp i \int_{w_0}^w P}.
$$

Note that using these operators we can obtain a given function
(for $\Sigma$ of genus zero)
$$
\Phi(z) = c + \sum_{i=1}^n \log(z-w^+_i) - \sum_{i=1}^n \log(z-w^-_i)
$$
as the correlator
\begin{equation}    \label{create}
\Phi(z) = \langle X(z) \prod_{i=1}^n \Psi_+(w^+_i) \prod_{i=1}^n
\Psi_-(w^-_i)  \left(\delta^2(X(z_0)-c_0) \psi(z_0) \ol\psi(\ol{z}_0)
\right) \rangle.
\end{equation}
The factor in brackets is needed so as to normalize the function
$\Phi(z)$ by the condition that $\Phi(z_0) = c_0$. This condition
naturally appears upon integrating over the zero modes of $X$ and
$\psi$.

Therefore we can {\em create} any meromorphic function on $\Sigma$ of
genus $0$ by taking the correlation function of the form
\eqref{create}. How to generalize this to $\Sigma$ of genus greater
than zero? In this case, for a meromorphic function to exist, the
points $w^\pm_i$ where it has zeroes and poles must satisfy a
constraint: the divisor $\sum_i (w_i^+) - \sum_i (w_i^-)$ has to be in
the kernel of the Abel-Jacobi map. Therefore for our theory to be
consistent, the correlation functions must somehow take this condition
into account.

This appears puzzling at first, but the apparent paradox is resolved
if we recall formula \eqref{P}. The one-form $P$ is defined up to an
addition of a linear combination of closed one-forms $\omega_j$, and
periodicity of the field $\phi$ implies that the coefficients $a_j$ in
front of these one-forms must be integer multiples $m_j$'s of
$2\pi$. In the path integral we need to sum up over the $m_j$'s, and
this leads to non-trivial consequences.

Let $\OO_i, i=1,\ldots,N$, be some local operators in the toric sigma
model and suppose we wish to compute the correlation function of these
as well as the holomortex operators $e^{- i \int^{w_j^+} P},
j=1,\ldots,n_+$, and $e^{i \int^{w_j^-} P}, j=1,\ldots,n_-$.  First of
all, recall from \secref{holomortex} that the number of insertions of
$e^{- i \int P}$ has to be equal to the number of insertions of $e^{i
\int P}$; otherwise, the correlation function is automatically
zero. Thus, $n_+ = n_- = n$. The correlation function should be a
differential form on the moduli space ${\mathcal M}_{g,N+2n}$. To
simplify our analysis, let us fix the complex structure on $\Sigma$
and the positions of the operators $\OO_i, i=1,\ldots,N$, leaving the
positions $w_j^\pm$ of the holomortex operators free, but
distinct. Consider the resulting differential form $\omega$ on the
configuration space of $2n$ distinct points on $\Sigma$. Its degree is
determined by the fermionic charge conservation. Assume for simplicity
that the operators $\OO_i$ do not contain fermions. As in genus zero,
we need to insert an operator of the form $\delta^2(X(z_0)-c_0)
\psi(z_0) \ol\psi(\ol{z}_0)$ to take care of the zero mode of $X$ and
$\psi$. This means that in addition we need to insert $g$ operators
$\pi$ and $\ol\pi$, so that we should get a $(g,g)$--form on the
configuration space.

According to a general prescription of \cite{W:cs,Zwiebach} (see also
\secref{cor fn} above), this differential form is constructed as
follows. It is completely determined by its values on $g$ tangent
vectors of the form $\pa/\pa w_j^\pm$ and $g$ tangent vectors of the
form $\pa/\pa \ol{w}_j^\pm$. Then at the corresponding point we have
to insert the operators $G_{-1} \Psi_\pm$, $\ol{G}_{-1} \Psi_\pm$ or
$G_{-1} \ol{G}_{-1} \Psi_\pm$. For example, let $j$ run from $1$ to
$g$. Then at the points $w_j^+$ we have to insert the operator $e^{ -
i \int^{w^+_j} P} \pi(w^+_j) \ol\pi(\ol{w}^+_j)$. The corresponding
value of our differential form $\omega$ is given by
$$
\langle \prod_{i=1}^N \OO_i \prod_{j=1}^g e^{- i \int^{w_j^+} P}
\pi(w^+_j) \ol\pi(\ol{w}^+_j) \prod_{j=g+1}^n e^{- i \int^{w_j^+} P}
\prod_{j=1}^n e^{i \int^{w_j^-} P} \rangle
$$
$$
= \langle \prod_{i=1}^N \OO_i \prod_{j=1}^g
\pi(w^+_j) \ol\pi(\ol{w}^+_j) \prod_{j=1}^n e^{i \int_{w_j^+}^{w_j^-}
  P} \rangle.
$$
Substituting formula \eqref{P}, we find that the correlation function
will contain the factor
$$
\exp \left( 2 \pi i \sum_k \sum_{j=1}^n m_k \int_{w_j^+}^{w_j^-}
\omega_k \right),
$$
and this is the only term that depends on the $m_k$'s. In the path
integral we will have to take the sum over all values of the
$m_k$'s. The result of this summation is a delta-function, which means
that the correlation function is identically equal to zero unless
$$
\sum_{j=1}^n m_k \int_{w_j^+}^{w_j^-} \omega_k = 0
$$
for all $k$. This precisely means that the divisor $\sum_j (w_j^+) -
\sum_j (w_j^-)$ has to be in the kernel of the Abel-Jacobi map.

Now it is clear that the differential form $\omega$ on the
configuration space of $2n$ points that we obtain in our theory is the
delta-form supported on the kernel of the Abel-Jacobi map (which has
codimension $g$). We can ``smoothen'' this delta-form by deforming the
action of our model with the term $\epsilon \int_\Sigma p \ol{p} d^2
z$ (see below).

Now suppose that $\OO_i, i=1,\ldots,N$, are operators from the sigma
model with the target $\pone$, and we wish to compute the correlation
function in the sigma model with the target $\pone$
\begin{equation}    \label{q-expansion}
\langle {\mc O}_1 \ldots {\mc O}_N \rangle_{\pone} = \sum_{n \geq 0}
\langle {\mc O}_1 \ldots {\mc O}_N \rangle_{\pone,n} q^n,
\end{equation}
where $\langle {\mc O}_1 \ldots {\mc O}_N \rangle_{\pone,n}$ is the
term corresponding to the holomorphic maps of degree $n$. As we
explained above, more general correlation functions in our sigma model
are obtained by inserting contour integrals of the fields $G(z)$ and
$\ol{G}(\ol{z})$ coupled to vector fields on ${\mathcal M}_{g,n}$.
These correlation functions are interpreted as differential forms on
${\mathcal M}_{g,n}$.

As we discussed above, in the setting of the linear sigma model with
the target $\C^\times$ the maps of degree $n$ are maps with
logarithmic singularities at $2n$ points $w_j^\pm, j=1,\ldots,n$
(counted with multiplicity). For fixed positions of these points such
a map is counted by inserting in the correlation function the
holomortex operators $\Psi_\pm(w_j^\pm,\ol{w}_j^\pm)$. Including {\em
all} possible positions of the points $w_j^\pm$ means applying to each
field $\Psi_\pm(w_j^\pm,\ol{w}_j^\pm)$ the operator $G_{-1}
\ol{G}_{-1}$, where $G_{-1}$ and $\ol{G}_{-1}$ are the contour
integrals of the fields $G(z)$ and $\ol{G}(\ol{z})$, coupled to the
translation vector field $\pa/\pa w_j^\pm$ and $\pa/\pa \ol{w}_j^\pm$,
respectively. In other words, we must replace each field
$\Psi_\pm(w,\ol{w})$ by the corresponding $(1,1)$--form
$\Psi_\pm^{(2)}(w,\ol{w})$ given by the formula
\begin{equation*}
\Psi_\pm^{(2)}(w,\ol{w}) = G_{-1} \ol{G}_{-1} \cdot
\Psi_\pm(w,\ol{w}) dw d \ol{w} = e^{ \mp i \int^w P} \pi(w)
\ol\pi(\ol{w}) dw d \ol{w},
\end{equation*}
and integrate these $(1,1)$--forms over $\Sigma$. Note that since the
operators $\Psi_\pm$ are $Q$--closed, $\Psi_\pm^{(2)}$ is the operator
obtained by cohomological descent (see \secref{warmup}).

Thus, we find that
\begin{multline*}
\langle {\mc O}_1(z_1,\ol{z}_1) \ldots {\mc O}_N(z_N,\ol{z}_N)
\rangle_{\pone,n} = \frac{1}{(n!)^2} \langle {\mc O}_1(z_1,\ol{z}_1)
\ldots {\mc O}_N(z_N,\ol{z}_N) \\ \times \prod_{i=1}^n \int_\Sigma
\Psi_+^{(2)}(w_j^+,\ol{w}_j^+) \prod_{i=1}^n \int_\Sigma
\Psi_-^{(2)}(w_j^-,\ol{w}_j^-) \rangle_{\C^\times}
\end{multline*}
(the coefficient $1/(n!)^2$ is due to the fact that the collections of
points $\{ w^+_j \}$ and $\{ w^+_j \}$ are unordered).

The integrand is not well-defined on the diagonals $w^+_i = w^-_j$
near the points $z_k$, a typical singularity being
$|z_k-w^+_i|^2/|z_k-w^j_-|^2$. However, we believe that the above
integrals do converge as long as we choose smooth observables $\OO_i$
(see an example in \secref{sample}). A proper way of treating this
integral may be to extend it to a compactification of ${\mathcal
M}_{g,N+2n}$. Note that the first $N$ points $z_1,\ldots,z_N$
correspond to the positions of the operators, while the additional
$2n$ points $w_1^\pm,\ldots,w_n^\pm$ correspond to parameters of the
space of maps $\Sigma \to \pone$. Therefore it is natural to expect
that the resulting compactification is related to the Kontsevich
moduli space of stable maps.

It follows from that we can write the correlation function $\langle
{\mc O}_1 \ldots {\mc O}_N \rangle_{\pone}$ as
\begin{multline*}
\langle {\mc O}_1 \ldots {\mc O}_N \rangle_{\pone} \\ = \langle {\mc
O}_1 \ldots {\mc O}_N \exp \left( q^{1/2} \int_\Sigma (\Psi_+(w)
\pi(w) \ol{\pi}(\ol{w}) + \Psi_-(w) \pi(w) \ol{\pi}(\ol{w})) dw
d\ol{w}\right) \rangle_{\C^\times}.
\end{multline*}
Therefore we have interpreted the correlation functions of the $\pone$
sigma model in the infinite volume as the correlation functions of
the deformation of the toric sigma model, with the deformed action
\begin{multline}    \label{action deformed}
\frac{i}{2\pi} \int_{\Sigma} d^2 z \; \left( p \pa_{\ol{z}} X +
\pi \pa_{\ol{z}} \psi + \ol{p} \pa_z \ol{X} + \ol\pi \pa_z \ol\psi
\right)
\\ + q^{1/2} \int_\Sigma (\Psi_+(w) \pi(w) \ol{\pi}(\ol{w}) +
\Psi_-(w) \pi(w) \ol{\pi}(\ol{w})) d^2 z.
\end{multline}
It is in this sense that we can say that the model with the deformed
action \eqref{action deformed} is equivalent to the type A twisted
sigma model with the target $\pone$ in the infinite volume.

This works fine when $\Sigma$ has genus zero. But for $\Sigma$ of
genus greater than zero, as we discussed in \secref{quantization}, we
need to take care of the zero modes of $p$ and $\ol{p}$. As we saw
above, the existence of these zero modes leads to correlation
functions being delta-like differential forms on the moduli spaces of
pointed curves. We can regularize these forms by adding the term
$\epsilon \int_\Sigma p \ol{p} d^2 z$ to the action. Note that we are
not adding the term corresponding to the inverse of the Fubini-Study
form on the target $\pone$, which would have violated conformal
invariance of the action, but rather the inverse of the {\em flat}
metric on $\C^\times$. While this flat metric has poles at $0,\infty
\in \pone$, its inverse has zeroes, and so it is regular on
$\pone$. This term preserves conformal invariance of our theory. There
is a similar regularization procedure in the case of more general Fano
toric varieties.

This regularization procedure becomes particularly important
for maps of low degrees, where without regularization it may be
impossible to evaluate the correlation functions.

To illustrate this point, consider the simplest example. Suppose that
$\Sigma$ is the torus and we wish to compute a contribution to some
correlation function corresponding to maps of degree one to
$\pone$. While there are certainly no maps from a smooth curve of
genus one to $\pone$, there are stable maps corresponding to curves
with nodal singularities having a genus zero component (this is often
referred to as ``bubbling''). Such maps constitute the entire moduli
space of stable maps in this case (unlike the case of maps of high
degree, where nodal curves contribute points at the boundary of the
locus corresponding to smooth curves). It is well-known that the
two-point function of the local observables ${\mc O}_1, {\mc O}_2$
corresponding to two-forms $\omega_1,\omega_2$ on $\pone$ such that
$\int \omega_i = 1$ is equal to $2q$ in this case. If we were to
follow the above recipe for the computation of the two-point function
in our deformed model literally, we would have to compute a
correlation function of the form
$$
q\langle {\mc O}_1(z_1,\ol{z}_1) {\mc O}_2(z_2,\ol{z}_2)
\int_\Sigma \Psi_+^{(2)}(w^+,\ol{w}^+) dw^+ d \ol{w}^+ \int_\Sigma
\Psi_-^{(2)}(w^-,\ol{w}^-) dw^- d \ol{w}^- \rangle_{\C^\times}.
$$
But as we explained above, the integral will be over those points
$w^+$ and $w^-$ which satisfy the Abel-Jacobi condition, which in this
case reads $w^+=w^-$. Since $$\Psi_+^{(2)}(w^+,\ol{w}^+)
\Psi_-^{(2)}(w^-,\ol{w}^-) \to 0$$ as $w^+ \to w^-$, it seems that we
obtain $0$.

However, if we deform the action by the term $\epsilon \int_\Sigma p
\ol{p} d^2 z$, the Abel-Jacobi condition is relaxed, and we obtain a
non-trivial integral. We will show elsewhere that this integral
reproduces the right answer $2q$ when $\epsilon \to 0$. We hope that
this is the mechanism by which we can ``reach'' the components of the
moduli spaces of stable maps which cannot be found in the closure of
the locus corresponding to smooth curves.

\ssec{Dual description of the deformed theory} \hfill    \label{dual}

\bigskip

\newenvironment{Cyr}{\fontencoding{OT2}\fontfamily{cmr}\selectfont}{}

\begin{Cyr}
\noindent \hspace*{90mm} A i B sideli na trube.\\
\hspace*{90mm} A upalo, B propalo.\\
\hspace*{90mm} Kto ostalsya na trube?
\end{Cyr}\footnote{{\em A and B were sitting on a pipe. A
fell, B disappeared. Who remained on the pipe?} (Russian folklore
riddle) The answer is ``and'', which is ``i'' in Russian; hence the
name ``I--model''.}

\bigskip

We have come to the key point of our construction. Let us apply the
$T$--duality of \secref{T-duality} to the deformed theory defined by
the action \eqref{action deformed}. In the dual variables $R, U$ the
holomortex operators $\Psi_\pm$ become purely local operators $e^{\pm
i U}$ and so the action \eqref{action deformed} becomes
\begin{equation}    \label{i model1}
\frac{i}{2\pi} \int_{\Sigma} d^2 z \; \left( \pa_z U \pa_{\ol{z}} R +
\pa_{\ol{z}} U \pa_z R + \pi \pa_{\ol{z}} \psi + \ol\pi \pa_z \psi
\right) + q^{1/2} \int_\Sigma (e^{iU} + e^{-iU}) \pi \ol{\pi} d^2 z.
\end{equation}
As we explained in the Introduction, this action is very similar to
the action \eqref{LG} of the B twisted Landau-Ginzburg model with the
superpotential $W(Y) = q^{1/2}(e^{iY} + e^{-iY})$.

Unlike the Lagrangian in \eqref{action deformed}, the Lagrangian in
\eqref{i model1} is local. The equivalence of the two theories implies
that the $q$--series expansion of the instanton contributions on the
deformed model described by \eqref{action deformed}, such as one given
by formula \eqref{q-expansion}, now has {\em non-perturbative} meaning
in the dual theory defined by \eqref{i model1}. In this theory
$q^{1/2}$ appears as the coupling constant, and if it is small, then
expanding the correlation functions in $q^{1/2}$ we reproduce the
$q$--expansion of the correlation functions of the sigma
model. However, we can study the theory with the action \eqref{i
model1} for arbitrary values of $q^{1/2}$.

Note that in the path integral definition of the correlation functions
of this model we must integrate over the single-valued function $U_0$
as well as over the integers $m_j = a_j/2\pi$ appearing in formula
\eqref{P}. This leads to some non-trivial consequences. In particular,
when $\Sigma$ has genus greater than zero, the correlation functions
involving the factor
$$
\prod_{j=1}^n e^{- i U(w_j^+)} \prod_{j=1}^n e^{i U(w_j^-)}
$$
are non-zero only if the divisor $\sum_j (w_j^+) - \sum_j (w_j^-)$ is
in the kernel of the Abel-Jacobi map. This follows in the same way as
for the toric sigma model (see \secref{deformation}).

Thus, the action \eqref{i model1} defines an intermediate model, which
we call the I--{\em model}, between the A--model, namely, the twisted
sigma model with the target $\pone$ in the infinite volume, and the
B--model, namely, the twisted Landau-Ginzburg model with the action
\eqref{LG}.

By the $T$--duality of \secref{T-duality}, the $q$--perturbative
I--model is equivalent to the A--model as a conformal field theory. On
the other hand, the correlation functions in the BPS sector of the
I--model are related to the correlation functions in the BPS sector of
the B--model, which is the Landau-Ginzburg model with the
superpotential $W$, considered in \cite{HV}, up to {\em contact terms}
(in the sense discussed in \cite{LNS}). Thus, we conclude that the
correlation functions in the BPS sector of the A--model are related to
the correlation functions in the BPS sector of the B--model
Landau-Ginzburg model, up to contact terms.  This is usually
considered as the statement of mirror symmetry.

Mathematically, this is expressed as the equality of certain
generating functions of Gromov-Witten invariants of $M$ (these
corresponding to correlation functions in the sigma model deformed by
the gravitational descendants) and certain oscillating integrals
(these correspond to the correlation functions in a Landau-Ginzburg
model). In general, this equivalence involves an intricate
transformation on the space of coupling constants that is referred to
as the mirror map (see \cite{Giv}, the recent book \cite{Mirror} and
references therein for details). The reason for this transformation is
that the two theories differ by contact terms, and this difference has
to be absorbed in a transformation of the coupling constants (see
\cite{LNS}).

To summarize, our construction for $M=\pone$ (and for the more general
case of a Fano toric variety $M$ treated in the next section) realizes
this correspondence of BPS correlation functions in two steps. First,
we have an equivalence of two conformal field theories, the twisted
sigma model of $M$ (A--model) and the intermediate model defined by
the action \eqref{i model1} (I--model). This means that all
correlation functions that one can write in the A--model and the
I--model are equal to each other. Second, we have a correspondence
between the I--model to the B--model, which is more subtle: it applies
only to the BPS sector, and in the BPS sector the two models are
equivalent only up to contact terms, which is the reason for
non-triviality of the mirror map. We do not address here the issue of
computing these contact terms and explicitly deriving the mirror map
from our proposed equivalence. But in principle this can be done. We
hope to return to this issue in a future paper.

\ssec{The supersymmetry charges}    \label{supercharges}

Recall that in the toric sigma model the left and right moving
supersymmetry charges are given by the formulas
$$
Q = - i \int \psi(z) p(z) dz, \qquad \ol{Q} = - i \int \ol\psi(\ol{z})
\ol{p}(\ol{z}) d\ol{z}.
$$
The total supersymmetry charge $Q + \ol{Q}$ corresponds to the de Rham
differential, which is typical for an A--model.


After the deformation to the theory with the action \eqref{action
deformed} the supercharges change their form. This is due to the fact
that the field $Q(z) = \psi(z) p(z)$ is no longer holomorphic and the
field $\ol{Q}(\ol{z})$ is no longer anti-holomorphic in the deformed
theory. In fact, for any chiral field $A(z)$ in a conformal field
theory, after deforming the action with the term $\int \Phi(z,\ol{z})
dz d\ol{z}$, we have the following formula (see \cite{Zam}):
\begin{equation}    \label{zam}
(\pa_{\ol{z}} A)(z,\ol{z}) = \int \Phi(w,\ol{z}) dw \cdot A(z),
\end{equation}
where the integral is over a small contour enclosing $z$. There is a
similar formula for an anti-chiral field.

Suppose that we have a superconformal field theory such that $\Phi =
\Psi^{(2)} = G_{-1} \ol{G}_{-1} \Psi$, where $\Psi$ is even,
$Q$--closed and a highest weight vector of the Virasoro algebra,
i.e., $L_n \Psi = \ol{L}_n \Psi = 0, n \geq 0$. Let $\Psi^{(1)}$ be
the one-form obtained by cohomological descent (see \secref{warmup}):
$$
\Psi^{(1)} = \Psi^{(1)}_z dz + \Psi^{(1)}_{\ol{z}} d \ol{z} = G_{-1}
\Psi dz + \ol{G}_{-1} \Psi d \ol{z}.
$$
Then if $A(z) = Q(z)$ we find that $$\int \Psi^{(2)} dw \cdot Q(z) = -
\pa_z \Psi^{(1)}_{\ol{z}}.$$ Hence the new left
moving supercharge is $\int (Q dz - \Psi^{(1)}_{\ol{z}}
d\ol{z})$. Likewise, we have
$$
\int \Psi^{(2)} d\ol{w} \cdot \ol{Q}(\ol{z}) = - \pa_{\ol{z}}
\Psi^{(1)}_{\ol{z}},
$$
and so the new right moving supercharge is $\int (\ol{Q}
d\ol{z} - \Psi^{(1)}_z dz)$. Thus, the total supercharge of the
deformed theory is
$$
Q + \ol{Q} - \int \Psi^{(1)}.
$$

In our case we have a deformation by $$\int q^{1/2} \left( \Psi^{(2)}_+
+ \Psi^{(2)}_- \right) dz d\ol{z},$$ where $\Psi^{(2)}_\pm = e^{\mp i
\int P} \pi \ol\pi$. Therefore we find that
$$
\Psi^{(1)}_{\pm,\ol{z}} = \pm i e^{\mp i \int P} \ol\pi.
$$
Thus, the new supercharge is
$$
Q(q) = - i \int \left( \psi p dz - q^{1/2} \left( e^{i \int P} -
  e^{- i \int P} \right) \ol\pi d\ol{z} \right).
$$
Similarly, we obtain that after the deformation the supercharge
$\ol{Q}$ becomes
$$
\ol{Q}(q) = - i \int \left( \ol\psi \ol{p} d\ol{z} + q^{1/2} \int
\left( e^{i \int P} - e^{- i \int P} \right) \pi dz \right).
$$

In the I--model, these supercharges look as follows:
\begin{align*}
Q(q) &= - i \int \left( \psi \pa_z U dz - q^{1/2} \left( e^{i U} -
e^{-i U} \right) \ol\pi d\ol{z} \right), \\ \ol{Q}(q) &= -i \int
\left( \ol\psi \pa_{\ol{z}} U d\ol{z} + q^{1/2} \left( e^{i U} - e^{-i
U} \right) \pi dz \right),
\end{align*}

Let us compute the {\em cohomology of the right moving} supercharge
$\ol{Q}(q)$ on the Hilbert space of our theory. This Hilbert space is
defined in \secref{T-dual, operator}. We will show in \secref{coh
pone} that the cohomology of the resulting complex coincides with the
the cohomology of a complex considered by L. Borisov \cite{Bor} and
F. Malikov and V. Schechtman in \cite{MS}. Its cohomology was shown in
\cite{MS} to be equal to the {\em quantum cohomology} of $\pone$. The
corresponding cohomology classes may be represented by $1$ and $e^{i
U} + e^{-i U}$.

On the other hand, according to \cite{W:new,Kapustin}, the cohomology
of the operator $\ol{Q}(q)$ in the {\em perturbative} regime (without
instanton corrections) should coincide with the cohomology of the
chiral de Rham complex of $\pone$. To obtain this result, we need to
consider a certain degeneration of the above complex, which
corresponds to the perturbative regime of the theory. For that we
introduce two parameters $t_1, t_2$ such that $t_1 t_2 = q$, and write
$t_1 e^{i U} - t_2 e^{- i U}$ instead of $q^{1/2}(e^{i U} - e^{-i
U})$. In the perturbative regime we have $t_1, t_2 \neq 0$, but their
product, which is $q$, becomes equal to $0$. In other words, we should
work over $\C[t_1,t_2]/(t_1 t_2)$. This corresponds to allowing only
degree zero maps $\Sigma \to \pone$. Such maps can pass through $0$ or
$\infty$, but not through both of them.

We will show in \secref{coh pone} that the cohomology of the
degenerate complex coincides with the cohomology of a complex
introduced in \cite{Bor} (see also \cite{MS}). Borisov showed in
\cite{Bor} that its cohomology is precisely the cohomology of the
chiral de Rham complex of $\pone$. Therefore we find an agreement with
the prediction of \cite{W:new,Kapustin}. Our computation explains the
meaning of the somewhat mysterious computation of \cite{Bor,MS} from
the point of view of the sigma model, with and without instanton
corrections.

\ssec{A sample computation of correlation functions}    \label{sample}

Here we show how to reproduce the simplest one-instanton calculation
of the A--model (the sigma model with the target $\pone$) in the
framework of the I--model defined by action \eqref{i model1}.

Let $\omega_i, i=1,2,3$, be three two-forms on $\pone$ representing
the second cohomology class. We will assume that they are invariant
under the $U(1)$--action on $\pone$ with the fixed points $0$ and
$\infty$. We identify $\pone \bs \{ 0,\infty \}$ with $\C/2\pi i\Z$
via the exponential map and use the coordinates $R$ and $\phi$ on
$\C/2\pi i\Z = \R \times i(\R/2\pi\Z)$ as before. With respect to
these coordinates, these forms may be written as $\omega_i = f_i(R) dX
d\ol{X}$, where $X=R+i\phi$. The local operators corresponding to the
two-forms $\omega_i$ in the A--model are
$$
\wh\omega_i = f_i(R) \psi \ol\psi.
$$

Consider the case when the worldsheet $\Sigma$ has genus zero. The
simplest non-trivial correlation function in the A--model is
$$
\langle \wh\omega_1 \wh\omega_2 \wh\omega_3 \rangle_{\pone} = q
\prod_{i=1}^3 \int_{\pone} \omega_i.
$$
Let us show how to reproduce this answer in the I--model.

In the I--model the operators $\wh\omega_i$ are given by the same
formula as above (since $R$ makes sense in the dual theory), hence
their correlation function expanded in powers of $q$ is the
correlation of the free field theory defined by the action
\eqref{total action} given by the formula
\begin{multline}    \label{instanton sum}
\langle \wh\omega_1 \wh\omega_2 \wh\omega_3 \exp\left( q^{1/2} \int
(e^{iU} + e^{-iU}) \pi \ol\pi \right) \rangle \\ = \sum_{n=0}^\infty
\frac{q^n}{(n!)^2} \langle \wh\omega_1 \wh\omega_2 \wh\omega_3 \left(
\int e^{iU} \pi \ol\pi \right)^n \left( \int e^{-iU} \pi \ol\pi
\right)^n \rangle.
\end{multline}
We have already explained above that, due to the charge conservation,
for the correlation function to be non-zero the number of insertions
of $e^{iU}$ has to be equal to the number of insertions of
$e^{-iU}$. This explains why in the above formula we consider only the
contributions corresponding to equal numbers of insertions.

Next, we count the ghost number. The chiral ghost number of each of
the operators $\wh\omega_i$ is one, due to the presence of the fermion
$\psi$. Hence the contribution of the operators $\wh\omega_i$ to the
chiral ghost number is $3$, and likewise for the anti-chiral ghost
number. The conservation law in genus zero is that the total chiral
number and the anti-chiral ghost number should be equal to $1$. Hence
to get a non-zero correlation function we must insert two chiral
fermions $\pi$ and two anti-chiral fermions $\ol\pi$. This means that
the only non-zero term in the sum \eqref{instanton sum} is the term
with $n=1$, and the coefficient in front of it is precisely $q$.

Thus, it remains to show that in the free field theory with the action
\eqref{total action} we have
\begin{equation}    \label{example}
\langle \prod_{i=1}^3 \wh\omega_i(z_i,\ol{z}_i) \int e^{iU} \pi \ol\pi
dw^- d \ol{w}^- \int e^{-iU} \pi \ol\pi dw^+ d\ol{w}^+ \rangle =
\prod_{i=1}^3 \int_{\pone} \omega_i.
\end{equation}
In the correlation function appearing in the left hand side of this
formula we have fixed the points $z_1,z_2,z_3$ and we are integrating
over the points $w^-$ and $w^+$ the $(1,1)$--forms $G_{-1} \ol{G}_{-1}
\cdot e^{\pm i U}$. By using the Ward identities in the standard way
(see \cite{W:cs,Zwiebach}), we can ``swap'' the operators $G_{-1}
\ol{G}_{-1}$ and the integrals from the variables $w^-$ and $w^+$ to
any two of the three variables $z_1,z_2,z_3$, say, $z_1$ and $z_2$,
fix the position of the remaining point $z_3$, say $z_3=\infty$, and
fix the positions of $w^-, w^+$. We find that
$$
G_{-1} \ol{G}_{-1} \cdot \wh\omega_i = f_i(R) \pa_z R \pa_{\ol{z}} R
dz d\ol{z}.
$$

The fermionic part of the correlation function becomes equal to $1$,
and the bosonic part is given by the integral
\begin{multline}    \label{triple}
\int d^2 X d^2 z_1 d^2 z_2 \langle f_1(R(z_1,\ol{z}_1))
f_2(R(z_2,\ol{z}_2)) f_3(R(\ol{z}_3)) \\ \times
\pa_{z_1} R(z_1) \pa_{\ol{z}_1} R(\ol{z}_1) \pa_{z_2} R(z_2)
\pa_{\ol{z}_2} R(\ol{z}_2) e^{iU(w^-)} e^{-iU(w^+)} \rangle
\end{multline}
(the integral over $d^2 X$ is the integral over the zero mode). But we
have the following OPE:
$$
R(z,\ol{z}) e^{\mp iU(w^\pm)} \sim \pm \log|z-w^\pm| e^{\mp iU(w^\pm)}.
$$
Hence
$$
\pa_{z} R(z) \pa_{\ol{z}} R(\ol{z}) e^{\mp iU(w^\pm)} \sim 
|z-w^\pm|^{\pm 2} e^{\mp iU(w^\pm)}.
$$
Therefore the term $\pa_{z} R(z) \pa_{\ol{z}} R(\ol{z})$ in the
correlation function \eqref{triple} may be replaced by
$|(z-w^+)/(z-w^-)|$, which is the Jacobian of the map $z \mapsto \log
c (z-w^+)/(z-w^-)$. Thus, the integrals over $z_1$ and $z_2$
correspond to the integrals of $\omega_1$ and $\omega_2$ over $\pone$,
while the integral over the zero mode corresponds to the integral of
$\omega_3$. We find that the integral \eqref{triple} is equal to the
right hand side of \eqref{example}, as desired. Note that in this
computation we have in effect ``localized'' on the holomorphic maps
$\Sigma \to \pone$ corresponding to the meromorphic functions
$c(z-w^+)/(z-w^-)$, where $c$ is a scalar.

\section{General toric varieties}    \label{general toric}

\ssec{Recollections on toric varieties}    \label{recollections}

Let us recall the combinatorial data involved in the definition of
smooth compact toric varieties, following \cite{Batyrev} (see also
\cite{V}).

Let $\Lambda$ be a lattice of rank $d$ and $\check\Lambda$ be the dual
lattice. We set $\Lambda_{\R} = \Lambda \otimes_{\Z} \R$,
$\Lambda_{\C} = \Lambda \otimes_{\Z} \C$. For $k \geq 1$ a convex
subset $\sigma \subset \Lambda_{\R}$ is called a regular
$k$--dimensional cone if it is generated by a subset of a basis of
$\Lambda$, i.e.,
$$
\sigma = \R_{\geq 0} \langle v_i \rangle_{i=1}^k = \left \{
\sum_{i=1}^k a_i v_i \, \big| \, a_i \in \R_{\geq 0} \right\},
$$
where $\{v_1,\ldots,v_k \}$ is a subset of $\Lambda$ that can be
extended to a basis. The $0$--dimensional regular cone is by
definition the origin $0 \in \Lambda_{\R}$. A subcone $\sigma'$ of
$\sigma$ generated by a subset of $\{ v_i \}_{i=1}^k$ is called a face
of $\sigma$. In this case we use the notation $\sigma' < \sigma$.

A finite collection $S = \{ \sigma_i \}_{i=1}^m$ is called a
complete regular fan if the following conditions are satisfied:
\begin{enumerate}
\item if $\sigma \in S$ and $\sigma' < \sigma$, then $\sigma' \in
  S$;

\item if $\sigma, \sigma' \in S$, then $\sigma \cap \sigma' <
  \sigma$ and $\sigma \cap \sigma' < \sigma'$.

\item $\Lambda_{\R} = \bigcup_{i=1}^m \sigma_i$.
\end{enumerate}

For example, let $\Lambda$ be the $d$--dimensional lattice generated
by $v_1,\ldots,v_d$. Set $v_{d+1} = - \sum_{i=1}^d v_i$. For any subset
$I \subset \{ 1,\ldots,d+1 \}$, let $\sigma_I = \R_{\geq 0} \langle
v_j \rangle_{j \in I}$. Then $S(d) = \{ \sigma_I \}_{I \subset \{
  1,\ldots,d+1 \}}$ is a complete regular fan.

One associates a toric variety to a fan $S$ as follows. To each
cone $\sigma \in S$ we assign the dual cone in $\check\Lambda$,
$$
\check\sigma = \{ \check\lambda \in \check\Lambda \, | \, \langle
\check\lambda,v \rangle \geq 0, \forall v \in \sigma \},
$$
and the affine variety ${\mathbb A}_\sigma = \on{Spec}
\C[\check\sigma]$. It is clear that if $\sigma' < \sigma$, then we
have a natural inclusion ${\mathbb A}_\sigma \hookrightarrow {\mathbb
A}_{\sigma'}$. This allows us to glue the varieties ${\mathbb
A}_\sigma, \sigma \in S$, into a projective variety ${\mathbb
P}_S$, which is the toric variety associated to $S$.

For example, the variety associated to the fan $S(d)$ is the
projective variety ${\mathbb P}^d$.

In particular, we have an open dense subvariety of ${\mathbb
  P}_S$,
$$
\TT_S = {\mathbb A}_{\{ 0 \}} = \on{Spec} \C[\check\Lambda]
\simeq \on{Spec} \C[x_i^{\pm 1}]_{i=1}^d = (\C^\times)^d.
$$
Here $x_i, i=1,\ldots,d$, are coordinates on $\TT_S$
corresponding to a basis $\{ \check{e}_1,\ldots,\check{e}_d \}$ of
$\check\Lambda$ that is dual to a basis $\{ e_1,\ldots,e_d \}$ of
$\Lambda$ that we fix once and for all. Note that any element
$\check\lambda = \sum_{i=1}^d a_i \check{e}_i$ gives rise to a
monomial function $\prod_{i=1}^d x_i^{a_i}$ on $\TT_S$
which we denote by $f_{\check\lambda}$.

In a basis independent way we can say that $\TT_S$ is the algebraic
torus, whose lattices of characters $\TT_S \to \C^\times$ and
cocharacters $\C^\times \to \TT_S$ are canonically identified with
$\check\Lambda$ and $\Lambda$, respectively.

Let $\sigma(1),\ldots,\sigma(N)$ be the set of all one-dimensional
cones in $S$. Each such cone $\sigma(i)$ has a canonical generator
$v(i) \in \Lambda$ that can be completed to a basis of $\Lambda$. The
varieties ${\mathbb A}_{\sigma(i)}, i=1,\ldots,N$ provide a covering
of the toric variety ${\mathbb P}_S$ by open dense subsets. By
definition, the ring of functions on ${\mathbb A}_{\sigma(i)}$ is the
span of all monomials $f_{\check\lambda}$, where $\langle
\check\lambda,v(i) \rangle \geq 0$. The complement of $\TT_S$ in
${\mathbb A}_{\sigma(i)}$ is the divisor $C_i$ in the latter whose
ideal is the span of the monomials $f_{\check\lambda}$, where $\langle
\check\lambda,v(i) \rangle > 0$. It is clear that the closures
$\ol{C}_i$ of these divisors are the irreducible components of the
complement of $\TT_S$ in $\PP_S$.

For instance, in the case of $\PP^d$, the one-dimensional cones are
$\sigma(i) = \R_{\geq 0} v_i, i=1,\ldots,d+1$, and so $v(i) = v_i$.
Therefore the varieties ${\mathbb A}_{\sigma(i)}$ are the subvarieties
of $\PP^d$, where all but the $i$th homogeneous components are
non-zero. The divisor $C_i$ consists of points in which the $i$th
homogeneous component is equal to $0$.

\ssec{The toric sigma model}

Let us fix a smooth compact toric variety $\PP_S$ corresponding to a
fan $S$. We will assume that $\PP_S$ is a fano variety. In fact, our
construction can be applied to more general toric varieties; however,
in the case of toric varieties that are not Fano the connection
between the deformed model that we define below and the A--model of
$\PP_S$ is more subtle. We have indicated some of the underlying
reasons for this in \secref{deformation}.

The first step of our construction is to define the toric sigma model
with the target
$$
\TT_S \simeq (\C^\times)^d = \on{Spec} \C[x_i^{\pm 1}]_{i=1}^d.
$$
This model is just the tensor product of $d$ independent copies of the
toric sigma model of $C^\times$ described in \secref{toric sigma}. We
will use the logarithmic coordinates $X_i, i=1,\ldots,d$, on
$(\C^\times)^d \simeq \Lambda_{\C}/2\pi i \Lambda$, such that $x_i =
e^{X_i}$. Thus, we have the fields $X^i,p_i,\psi^i,\pi_i$ and their
complex conjugates $X^{\ol{i}}, p_{\ol{i}}, \psi^{\ol{i}},
\pi_{\ol{i}}$.

For any element $\check\la = \sum_{i=1}^d a_i \check{e}_i \in \cLa$ we
have fields $X^{\cla} = \sum_{i=1}^d a_i X^i$ and $\ol{X}^{\cla} =
\sum_{i=1}^d a_i X^{\ol{i}}$, whereas for any element $\la =
\sum_{i=1}^d b_i e_i \in \La$ we have fields $p_\la = \sum_{i=1}^d
b_i p_i$ and $\ol{p}_\la = \sum_{i=1}^d b_i p_{\ol{i}}$. We define the
fermions $\psi^{\cla}, \ol{\psi}^{\cla}, \cla \in \cLa$, and $\pi_\la,
\ol\pi_\la, \la \in \La$ in the same way.

The action of the toric sigma model is given by the formula
\begin{equation}    \label{action infinity1}
\frac{i}{2\pi} \int_{\Sigma} d^2 z \; \left( p_i \pa_{\ol{z}} X^i + \pi_i
\pa_{\ol{z}} \psi^i + p_{\ol{i}} \pa_z X^{\ol{i}} + \pi_{\ol{i}} \pa_z
\psi^{\ol{i}} \right).
\end{equation}

The theory has $N=(2,2)$ superconformal symmetry. The corresponding
generators are the sums of the generators in the $\C^\times$ toric
sigma model given by formula \eqref{N=2}.

As in the one-dimensional case, explained in \secref{holomortex}, we
find that the fields $X^i$ may have non-trivial winding. The winding
numbers take values in the lattice $\La$. For each $\la \in \La$ we
introduce the corresponding {\em holomortex operators}
$$
\Psi_{\la}(z,\ol{z}) = e^{- i \int P_\la} = \exp\left(- i \int_{z_0}^z
(p_\la(w)dw + \ol{p}_\la(\ol{w})d\ol{w}) \right).
$$
They have the following OPE with the fields $X^{\cmu}$ and
$\ol{X}^{\cmu}$:
\begin{align*}
X^{\cmu}(z) \Psi_{\la}(w,\ol{w}) &= \langle \cmu,\la \rangle \log(z-w)
\Psi_\la(w,\ol{w}), \\
\ol{X}^{\cmu}(z) \Psi_{\la}(w,\ol{w}) &= \langle \cmu,\la \rangle
\log(\ol{z}-\ol{w}) \Psi_\la(w,\ol{w}).
\end{align*}
The prescription for the computation of correlation functions of these
operators is the same as in the one-dimensional case (see
\secref{holomortex}).

Next, we define the $T$--dual theory of the $\TT_S$--toric sigma
model. This is an ordinary sigma model with the target being the {\em
partially dualized torus} $$\check{\TT}_S = \Lambda_{\R} \times i
(\cLa_{\R}/2\pi \check\Lambda),$$ equipped with the Minkowski metric,
which is the product of $d$ copies of the Minkowski metric introduced
in \secref{T-duality}. Note that this metric is canonically defined
precisely because the the lattices $\Lambda$ and $\cLa$ are dual to
each other.

In the dual theory the bosonic fields are $U_i$ and $R^i$,
$i=1,\ldots,d$, and the fermionic fields are the same as in the toric
sigma model. The action is as in \eqref{total action}:
\begin{equation}    \label{total action1}
\wt{I} = \frac{i}{2\pi} \int_\Sigma d^2 z \; (\pa_z U_j \pa_{\ol{z}} R^j +
\pa_{\ol{z}} U_j \pa_z R^j + \pi_j \pa_{\ol{z}} \psi^j + \pi_{\ol{j}}
\pa_z \psi^{\ol{j}}).
\end{equation}

The transformation formulas for the bosonic fields of the two models
are
$$
p_i(z) = \pa_z U_i(z,\ol{z}), \qquad p_{\ol{i}}(\ol{z}) = \pa_{\ol{z}}
U_i(z,\ol{z}),
$$
$$
\frac{1}{2}(X^i(z) + X^{\ol{i}}(\ol{z})) = R^i(z,\ol{z}).
$$
The holomortex operators $\Psi_\la = e^{- i \int P_\la}$ have a
simple realization in the dual variables:
$$
\Psi_\la(z,\ol{z}) = e^{- i U_\la(z,\ol{z})},
$$
where we set $U_\la = \sum_{i=1}^d b_i U_i$ for $\la = \sum_{i=1}^d
b_i e_i$.

\ssec{Changing the target from $\TT_S$ to $\PP_S$}

We wish to describe the non-linear sigma model with the target toric
variety $\PP_S$ as a deformation of the toric sigma model with the
target torus $\TT_S$. We follow the same idea as in the case of
$\pone$ explained in \secref{changing to pone}. Recall from
\secref{recollections} that the complement of $\TT_S$ in $\PP_S$ is a
divisor, whose irreducible components $\ol{C}_j, j=1,\ldots,N$, are
naturally parameterized by the one-dimensional cones $\sigma_j$ in
$S$ generated by $v(j) \in \Lambda$. A generic holomorphic map $\Phi:
\Sigma \to \PP_S$ takes values in $\TT_S \subset \PP_S$ for all but
finitely many points, and at the special points it takes values in the
open part $C_j$ of the divisor $\ol{C}_j$, introduced in
\secref{recollections}, for some $j=1,\ldots,N$. Let us denote the
points of $\Sigma$ where $\Phi$ takes values in $C_i$ by $w^{(j)}_k,
j=1,\ldots,m_j$.

We propose to {\em include such maps by inserting in the correlation
functions the holomortex operators}
$\Psi_{v(j)}(w^{(j)}_k,\ol{w}^{(j)}_k)$ introduced in the previous
section. Recall that
$$
\Psi_{v(j)}(w,\ol{w}) = e^{-i \int_{w_0}^w P_{v(j)}}.
$$
Clearly, these operators are $Q$--closed. Hence we find the following
formula for the two-form cohomological descendant field of
$\Psi_{v(j)}(w,\ol{w})$:
$$
\Psi^{(2)}_{v(j)}(w,\ol{w}) dw d\ol{w} = \Psi_{v(j)}(w,\ol{w})
\pi_{v(j)}(w) \ol\pi_{v(j)}(\ol{w}) dw d\ol{w}.
$$

Now observe that the lattice of all relations between the generators
$v(j), j=1,\ldots,N$, of one-dimen\-sional cones in $S$ is generated
by $N-d$ linearly independent relations
$$
\sum_{j=1}^N a_{ij} v(j) = 0, \qquad i=1,\ldots,N-d,
$$
where we choose the $a_{ij}$'s to be integers that are relatively
prime.

Let us introduce parameters $t_j, j=1,\ldots,N$, and set
\begin{equation}    \label{ti}
q_i = \prod_{j=1}^N t_j^{a_{ij}}, \qquad i=1,\ldots,N-d.
\end{equation}

As in the case of $\pone$, the type A twisted sigma model with the
target $\PP_S$ in the infinite volume is then described by the
deformation of the toric sigma model by
$$
\sum_{j=1}^N t_j \int_\Sigma \Psi_{v(j)} \pi_{v(j)} \ol\pi_{v(j)}
dw d\ol{w}.
$$
Note that the $t_j$'s can be redefined by changing the normalization
of the operators $\Psi_{v(j)}$, but this will not affect the
parameters $q_i$ given by formula \eqref{ti}. Therefore the $q_i$'s
are the true parameters of the theory, and they correspond precisely
to the K\"ahler classes on $\PP_S$, as explained in \cite{Batyrev}.

For example, if $\PP_S = \PP^d$, then we have
$$
\Psi_{v(j)} \pi_{v(j)} \ol\pi_{v(j)} = e^{-i \int P_j} \pi_j \ol\pi_j,
\qquad j=1,\ldots,d,
$$
$$
\Psi_{v(d+1)} \pi_{v(d+1)} \ol\pi_{v(d+1)} = e^{i \sum_{j=1}^d \int
  P_j} \left( \sum_{j=1}^d \pi_j \right) \left( \sum_{j=1}^d
\pi_{\ol{j}} \right),
$$
and there is only one parameter $q = \prod_{j=1}^{d+1} t_j$.

\ssec{The I--model}

Finally, we apply the $T$--duality to the action of the deformed toric
sigma model. The operators $\Psi_{v(j)}$ are now written as $e^{- i
  U_{v(j)}}$, and so the action takes the form
\begin{equation}    \label{i model2}
\frac{i}{2\pi} \int_{\Sigma} d^2 z \; \left( \pa_z U_j \pa_{\ol{z}} R^j +
\pa_{\ol{z}} U_j \pa_z R^j + \pi_j \pa_{\ol{z}} \psi^j + \pi_{\ol{j}}
\pa_z \psi^{\ol{j}} \right) + \int_\Sigma \wt{W} d^2 z,
\end{equation}
where
$$
\wt{W} = \sum_{j=1}^N t_j e^{- i U_{v(j)}} \pi_{v(j)}
  \ol\pi_{v(j)}.
$$
For example, if $\PP_S = \PP^d$, then we have
$$
\wt{W} = \sum_{j=1}^d t_j e^{- i U_j} \pi_j \pi_{\ol{j}}
+ t_{d+1} e^{i \sum_{j=1}^d U_j} \left( \sum_{j=1}^d \pi_j \right)
\left( \sum_{j=1}^d \pi_{\ol{j}} \right).
$$

The action \eqref{i model2} defines the I--model for a general toric
variety $\PP_S$. As in the case of $\pone$, the action \eqref{i
  model2} should be compared to the action of the Landau-Ginzburg
model with the superpotential
\begin{equation}    \label{superpotential}
W = \sum_{j=1}^N t_j e^{- i Y_{v(j)}}.
\end{equation}
Here $Y_{v(j)}$ is a chiral superfield which is a linear combination
of $d$ independent chiral superfields $Y_k, k =1,\ldots,d$, defined by
the formula $Y_{v(j)} = \sum_{k=1}^d b_{ij} Y_k$, where $v(j) =
\sum_{k=1}^d b_{ij} e_k$. We recognize in formula
\eqref{superpotential} the superpotential of the type B twisted
Landau-Ginzburg model that is mirror dual to the type A sigma model
with the target $\PP_S$ considered in \cite{HV}.

As we explained in the case of $\pone$, this suggests that mirror
symmetry can be realized in two steps. The first step is the
equivalence of the twisted sigma model of $\PP_S$ (A--model),
described as a deformation of a free field theory, and the
intermediate model defined by the action \eqref{i model2} (I--model),
as conformal field theories. The second step is a correspondence
between the I--model to the B--model, which is more subtle: it applies
only to the BPS sector, and in the BPS sector the two models are
equivalent only up to contact terms.

We hope that further study of the I--model and its connections with
the A--model on the one hand and the B-model on the other hand will
help us understand more fully the nature of mirror symmetry.

\ssec{Supercharges}    \label{supercharges1}

Let us compute the supersymmetry charges of the I--model. Following
the same computation as in \secref{supercharges}, we find the
following formulas for the left and right moving supercharges:
\begin{align*}
Q &= - i \int \left( \psi^k \pa_z U_k dz + \left( \sum_{j=1}^N t_j
e^{- i U_{v(j)}} \ol\pi_{v(j)} d\ol{z} \right) \right), \\ \ol{Q}
&= - i \int \left( \psi^{\ol{k}} \pa_{\ol{z}} U_k d\ol{z} - \left(
\sum_{j=1}^N t_j e^{- i U_{v(j)}} \pi_{v(j)} dz \right) \right).
\end{align*}

It is interesting to compute the cohomologies of the right moving
supercharge $\ol{Q}(q)$. As in the case of $\pone$, we will find in
\secref{cohomology} that these cohomologies coincide with the
cohomologies of a complex constructed in \cite{Bor,MS}. It was shown
in \cite{MS} that in the case of $\PP^n$ this cohomology coincides
with the quantum cohomology of $\PP^n$. We expect the same to be true
for more general toric Fano varieties. (In fact, it follows from the
results of \cite{MS} that the cohomology of the total supercharge $Q +
\ol{Q}$ is isomorphic to the quantum cohomology of $\PP_S$.) On the
other hand, the cohomology of a degeneration of this complex was
computed in \cite{Bor}, and it gives the cohomology of the chiral de
Rham complex of $\PP_S$. This agrees with the prediction of
\cite{W:new,Kapustin}.

\section{Operator formalism}    \label{operator}

In this section we discuss the operator content of the toric sigma
models introduced in the previous sections, the $T$--duality transform
and the deformed models. For simplicity we will mostly treat the case
of the target $\C^\times$ as the general case is very similar. The
algebraic object that we define (we call it the ``Hilbert space'' of
the theory) obeys the axioms of a vertex algebra mixing chiral and
anti-chiral sectors, similar to the ones defined by A. Kapustin and
D. Orlov in \cite{KO} who considered the case of sigma models of the
torii in the finite volume. In particular, we define a state-field
correspondence assigning to every state of the Hilbert space an
operator depending on $z,\ol{z}$ acting on the Hilbert space. Thus,
the toric sigma models and their $T$--dual models studied in this
paper provide us with new examples of vertex algebras in which chiral
and anti-chiral sectors are non-trivially mixed. While there is a vast
mathematical literature on the subject of chiral algebras, examples of
mixed vertex algebras have not been widely discussed in the
mathematical literature so far. Actually, it is expected that the
vertex algebras that occur in the study of mirror symmetry are for the
most part of this sort, with the chiral and anti-chiral sectors
entangled in a non-trivial way. Therefore we believe that algebraic
study of such vertex algebras is important.

At the end of this section we will compute the chiral algebra of the
I--model and the cohomology of the right moving supercharge, making a
connection with the results of \cite{Bor,MS}.

\subsection{Hilbert space and state-field correspondence in the toric
  sigma model}
\label{hamiltonian}

We collect all the ingredients found in \secref{toric sigma model} and
define the Hilbert space and the state-field correspondence of the
toric sigma model.

Let us write
\begin{align*}
X(z) &= \om \log z + \sum_{n \in \Z} X_n z^{-n}, \\ \ol{X}(\ol{z}) &=
\om \log \ol{z} + \sum_{n \in \Z} \ol{X}_n \ol{z}^{-n}, \\
p(z) &= \sum_{n \in \Z} p_n z^{-n-1}, \\ p(z) &= \sum_{n \in \Z}
\ol{p}_n \ol{z}^{-n-1},
\end{align*}
and let $T_m$ be the operator satisfying
$$
[\om,T_m] = m T_m
$$
and commuting with the $X_n$'s and $p_n$'s. Note that we also have the
following commutation relations:
\begin{align*}
[X_n,p_m] &= - i \delta_{n,-m}, \\ [p_n,p_m] &= [X_n,X_m] = 0,
\end{align*}
and $\om$ commutes with all $p_n$'s and $X_n$'s. We also have similar
formulas for the components of the anti-chiral fields.

Consider the Heisenberg algebras generated by $X_n,p_n, n \in \Z$, and
$\ol{X}_n,\ol{p}_n, n \in \Z$, respectively. For $\gamma \in \C$, let
$\F_{\gamma}$ (resp., $\ol\F_{\gamma}$) be the Fock representation of
the Heisenberg algebra generated by a vector annihilated by $X_n, n>0,
p_m, m \geq 0$ (resp., $\ol{X}_n, n>0, \ol{p}_m, m \geq 0$) and on
which $ip_0$ (resp., $i\ol{p}_0$) acts by multiplication by
$\gamma$. Since the imaginary part of $X(z)$ is periodic, the
eigenvalues of $i(p_0-\ol{p}_0)$ are quantized to be integers. This is
exactly the condition that we obtained in \secref{holomortex}. The
operator $\om$ is also quantized and has to take integer eigenvalues,
called the winding numbers.

The Fock representation $\F_\al$ (resp., $\ol\F_{\al}$) on which $\om$
acts by multiplication by $m \in \Z$ will be denoted $\F_{\al,m}$
(resp., $\ol\F_{\al,m}$). We denote by $|\al,n\rangle$ (resp.,
$\ol{|\al,m\rangle}$) its generating vector.

The big bosonic Hilbert space of the theory is the direct product of
the tensor products of the left and right moving Fock representations
\begin{equation}    \label{Fock spaces}
\F_{(r+\al)/2,m} \otimes \ol{\F}_{(-r+\al)/2,m},
\end{equation}
where $r, m \in \Z$ and $\al$ runs over a subset of $\C$. There are
different choices for this subset which is determined by what type of
functions of the zero mode $R_0$ of the field $R(z,\ol{z}) =
(X(z)+\ol{X}(\ol{z}))/2$ we wish to allow.

One possibility is to restrict ourselves to the subset of $\al \in i
\R \subset \C$. This is compatible with the structure of the Hilbert
space in the sigma model at the finite radius $\sqrt{t}$ and
corresponds to restricting ourselves to the $L_2$ functions of the
zero mode. This choice is natural from the point of view of the latter
model because it can itself be obtained as the sigma model with the
target torus of radii $\sqrt{t}$ and $r$ in the limit when $r \to
\infty$.

But this is not the only way to treat the toric sigma model. Indeed,
we will consider it as a degeneration of the sigma model with the
target $\pone$ (in the infinite volume limit). Therefore another
natural choice for the class of functions of the zero mode is the
space of polynomial functions in $e^{\pm R_0}$. In fact, it is natural
to consider all rational functions on $\pone$ which are regular on
$\C^\times$, that is polynomials in $e^{\pm X_0}$, as well as their
complex conjugates, polynomials in $e^{\pm \ol{X}_0}$. Choosing this
space is equivalent to demanding that $\al$ be in the set $\Z \subset
\C$. In the subsequent sections we will define a deformation of the
toric sigma model which is equivalent to the A--model of $\pone$ (that
is the type A twisted sigma model with the target $\pone$ in the
infinite volume). The operators in this theory will be obtained by
restriction to $\C^\times$ from operators defined on the entire
$\pone$. While there are no regular functions on $\pone$ other than
constant functions, there will be composite operators depending on
$X(z)$ as well as $p(z)$ that are well-defined, such as the normally
ordered products $\Wick e^{\pm X(z)} p(z) \Wick$.

Thus, we see that there are different choices for the subset of
$\al$'s which we may include in our Hilbert space. However, from the
purely algebraic point of view, the state-field correspondence that we
will now describe works equally well for any of these
choices. Therefore in the rest of this subsection we will consider the
direct product of the Fock spaces \eqref{Fock spaces} with arbitrary
complex values of $\al$.

The state-field correspondence that we describe now gives us the
structure of a vertex algebra combining holomorphic and
anti-holomorphic sections, in the sense of \cite{KO}. Note that just
like in the case of sigma models on the torii that was considered in
\cite{KO}, we cannot separate the holomorphic and anti-holomorphic
sectors of this vertex algebra.

The key point is the assignment of a field to the state $|0,m\rangle
\otimes \ol{|0,m\rangle}$. We assign to it the field $e^{- im\int P}$
given by the formula
\begin{multline}    \label{Phim}
e^{- i m\int P}(z,\ol{z}) = \exp \left( - i m \int^z (p(w) dw +
\ol{p}(\ol{w}) d\ol{w}) \right) \; \overset{\on{def}}= \\ T_{m} |z|^{-
im(p_0+\ol{p}_0)} \left( \frac{z}{\ol{z}} \right)^{-
im(p_0-\ol{p}_0)/2} \exp \left( im \sum_{n \neq 0} \frac{p_n z^{-n} +
\ol{p}_n \ol{z}^{-n}}{n} \right),
\end{multline}
where $T_m$ is the translation operator that shifts the winding number
by $m$ and commutes with all other operators. Note that the operator
$i(p_0-\ol{p}_0)$ has only integer eigenvalues on the Hilbert space of
the theory, so this formula is well-defined.

This field has the following OPE with $X(z)$:
$$
X(z) e^{- i m\int P}(w,\ol{w}) = m \log(z-w) e^{- i m\int P}(w,\ol{w})
,
$$
and similarly with $\ol{X}(\ol{z})$.

Next, we define the field corresponding to the state
$$
|(r+\al)/2,m\rangle \otimes \ol{|(-r+\al)/2,m\rangle}
$$
as the normally ordered product
$$
\Wick e^{(r+\al)X(z)/2+(-r+\al)\ol{X}(\ol{z})/2} e^{-im\int
P}(z,\ol{z})
\Wick\, ,
$$
where
\begin{multline*}
e^{(r+\al)X(z)/2+(-r+\al)\ol{X}(\ol{z})/2} \; \overset{\on{def}}= \\
|z|^{\om\al} \left( \frac{z}{\ol{z}} \right)^{r\om/2}
S_{(r+\al)/2,(-r+\al)/2} \exp \left( \frac{1}{2} (r+\al) \sum_{n \neq
  0} X_n z^{-n} + \frac{1}{2} (-r+\al) \sum_{n \neq 0} \ol{X}_n
\ol{z}^{-n} \right).
\end{multline*}
Here $S_{(r+\al)/2,(-r+\al)/2}$ is the translation operator
$$
\F_{(r'+\al')/2,m} \otimes \ol{\F}_{(-r'+\al')/2,m} \to
\F_{(r+r'+\al+\al')/2,m} \otimes \ol{\F}_{(-r-r'+\al+\al')/2,m}.
$$
Note that since $\om$ has only integer eigenvalues, this formula is
well-defined.

Finally, the fields corresponding to other states in the Fock
representation $\F_{r+\al,m} \otimes \ol{\F}_{-r+\al,m}$ are
constructed as the normally ordered products of the fields defined
above and the fields $\pa_z X(z), p(z), \pa_{\ol{z}} \ol{X}(\ol{z}),
\ol{p}(\ol{z})$, under the usual assignment:
$$
X_n \mapsto \frac{1}{(-n)!} \pa_z^{-n} X(z), \quad n\leq 0; \qquad p_n
\mapsto \frac{1}{(-n-1)!} \pa_z^{-n-1} p(z), \quad n<0.
$$
This completes the description of the bosonic part of the Hilbert
space of the theory.

Now we describe the fermionic part. Let us write
$$
\psi(z) = \sum_{n \in \Z} \psi_n z^{-n}, \qquad \pi(z) = \sum_{n \in
  \Z} \pi_n z^{-n-1},
$$
and similarly for the anti-chiral fields. The operator product
expansions give us the following anti-commutation relations:
$$
[\pi_n,\psi_m]_+ = -i \delta_{n,-m}, \qquad [\pi_n,\pi_m]_+ =
[\psi_n,\psi_m]_+ = 0,
$$
and similarly for the components of the anti-chiral fields. Consider
the Clifford algebra generated by $\psi_n, \pi_n, n \in \Z$ (resp.,
$\ol\psi_n, \ol\pi_n, n \in \Z$) and let $\F_{\on{ferm}}$ (resp.,
$\ol\F_{\on{ferm}}$) be the fermionic Fock space representation of
this algebra generated by a vector annihilated by $\psi_n, n>0, \pi_m,
m\geq 0$ (resp., $\ol\psi_n, n>0, \ol\pi_m, m\geq 0$). The fermionic
Hilbert space is $\F_{\on{ferm}} \otimes \ol\F_{\on{ferm}}$. The total
Hilbert space of the theory is the tensor product of the bosonic and
fermionic spaces.

\subsection{$T$--duality, operator formalism}    \label{T-dual,
  operator}

Next, we discuss the duality transformation from the operator point of
view. The operator content of the toric sigma model is described in
the previous section. Let is now describe the operator content of the
$T$--dual free bosonic field theory given by the action \eqref{bos
action}. The equations of motion imply that the fields $U$ and $R$ are
harmonic, so we can write
\begin{align*}
R(z,\ol{z}) &= R_0 + \log |z| p_R + \sum_{n \neq 0}
\frac{R_n}{n} z^{-n} + \sum_{n \neq 0}
\frac{\ol{R}_n}{n} \ol{z}^{-n}, \\
U(z,\ol{z}) &= U_0 + \log |z| p_U - \frac{i}{2} \wt{\om} \log
\frac{z}{\ol{z}} + \sum_{n \neq 0}
\frac{U_n}{n} z^{-n} + \sum_{n \neq 0}
\frac{\ol{U}_n}{n} \ol{z}^{-n}.
\end{align*}
The OPEs of these fields are of the form
$$
R(z,\ol{z}) U(w,\ol{w}) = i \log|z-w| + \Wick R(z,\ol{z}) U(w,\ol{w})
\Wick \, .
$$

The Fourier coefficients of these fields satisfy the following
commutation relations
$$
[R_n,U_m] = \frac{i}{2} n\delta_{n,-m}, \qquad [R_n,R_m] = [U_n,U_m]
= 0,
$$
$$
[\ol{R}_n,\ol{U}_m] = \frac{i}{2} n\delta_{n,-m}, \qquad
[\ol{R}_n,\ol{R}_m] = [\ol{U}_n,\ol{U}_m] = 0,
$$
and
$$
[R_0,p_U] = - i, \qquad [U_0,p_R] = i.
$$
All other commutators are equal to zero. Because $U$ is periodic, the
momentum operator $p_R$ is quantized and takes only integer values,
whereas $p_U$ can take arbitrary values. Also, the winding operator
$\wt\omega$ is quantized and takes only integer values.

The Hilbert space of the theory is built from Fock representations of
the Heisenberg algebra generated by the coefficients in the expansions
of $R$ and $U$. For $\beta \in \C$ and $r,m \in \Z$, let
$\wt\F_{r,\beta,m}$ be the Fock representation generated by a vector
$|r,\beta,m\rangle$ annihilated by $R_n, U_n, \ol{R}_n,
\ol{U}_n, n>0$, and on which $p_R$ and $p_U$ act by multiplication
by $m$ and by $\beta$, respectively, and $\wt{\om}$ acts by
multiplication by $r$.

Introduce the following translation operators, which map generating
vectors to generating vectors and commute with the operators $R_n,
U_n, \ol{R}_n, \ol{U}_n, n \neq 0$:
\begin{align*}
e^{\beta R_0}: \qquad &\wt\F_{r',\beta',m'} \to \wt\F_{r',\beta'+\beta,m'},
\\ e^{imU_0}: \qquad &\wt\F_{r',\beta',m'} \to \wt\F_{r',\beta,m'+m}, \\
e^{r \wh{R}_0}: \qquad &\wt\F_{r',\beta',m'} \to \wt\F_{r'+r,\beta',m'}.
\end{align*}

The state-field correspondence is defined as follows. The field
corresponding to the vector $|r,\beta,m \rangle$ is given by the
normally ordered product
\begin{equation}    \label{exp field}
\Wick e^{\beta R(z,\ol{z})} e^{i m U(z,\ol{z})} e^{r
\wh{R}(z,\ol{z})} \Wick \, ,
\end{equation}
where
\begin{align*}
e^{\beta R(z,\ol{z})} &= e^{\beta R_0} |z|^{\beta p_R} \Wick \exp
\left( \beta \sum_{n \neq 0} \frac{R_n}{n} z^{-n} + \beta \sum_{n
  \neq 0} \frac{\ol{R}_n}{n} \ol{z}^{-n} \right) \Wick \, , \\
e^{i m U(z,\ol{z})} &= e^{i m U_0} |z|^{i m p_U} \left( \frac{z}{\ol{z}}
\right)^{m\wt{\om}/2} \Wick \exp \left( i m \sum_{n \neq 0} \frac{U_n}{n}
z^{-n} + m \sum_{n \neq 0} \frac{\ol{U}_n}{n} \ol{z}^{-n} \right)
\Wick \, , \\
e^{r\wh{R}(z,\ol{z})} &= e^{r \wh{R}_0} \left(
\frac{z}{\ol{z}} \right)^{r p_R/2} \Wick \exp
\left( r \sum_{n \neq 0} \frac{R_n}{n} z^{-n} - r
\sum_{n \neq 0} \frac{\ol{R}_n}{n} \ol{z}^{-n} \right) \Wick \, .
\end{align*}
The other fields are obtained in the standard way as normally ordered
products of the field \eqref{exp field} and the derivatives of
$R$ and $U$, under the rule
$$
R_n \mapsto \frac{1}{(-n-1)!} \pa_z^{-n} R(z,\ol{z}), \qquad \ol{R}_n
\mapsto \frac{1}{(-n-1)!} \pa_{\ol{z}}^{-n} R(z,\ol{z}),
$$
$$
U_n \mapsto \frac{1}{(-n-1)!} \pa_z^{-n} U(z,\ol{z}), \qquad \ol{U}_n
\mapsto \frac{1}{(-n-1)!} \pa_{\ol{z}}^{-n} U(z,\ol{z}),
$$
for $n<0$.

The isomorphism between the Hilbert spaces of the two theories is
given by the following transformation of the generating fields:
\begin{equation}    \label{trans of fields 1}
\frac{1}{2}(X(z) + \ol{X}(\ol{z})) \mapsto R(z,\ol{z}),
\end{equation}
\begin{equation}    \label{trans of fields 2}
p(z) \mapsto \pa_z U(z,\ol{z}), \qquad \ol{p}(\ol{z}) \mapsto
\pa_{\ol{z}} U(z,\ol{z}),
\end{equation}
\begin{equation}    \label{trans of fields 3}
\frac{1}{2}(X(z) - \ol{X}(\ol{z})) \mapsto \wh{R}(z,\ol{z}).
\end{equation}
The field $\wh{R}(z,\ol{z})$ is non-local with respect to
$R(z,\ol{z})$, namely, $\wh{R}(z,\ol{z}) = R_-(z) - R_+(\ol{z})$,
where $R_\pm$ are the holomorphic and anti-holomorphic parts of
$R(z,\ol{z}) = R_-(z) + R_+(\ol{z})$ defined by the formulas
\begin{align}    \label{R-}
R_-(z) &= \frac{1}{2}(R^-_0 + p_R \log z) + \sum_{n \neq 0}
\frac{R_n}{n} z^{-n}, \\ \notag
R_+(\ol{z}) &= \frac{1}{2}(R^+_0 + p_R \log \ol{z}) + \sum_{n \neq 0}
\frac{\ol{R}_n}{n} \ol{z}^{-n},
\end{align}
where $R^\pm_0 = R_0 \mp \wh{R}_0$.

More precisely, at the level of the operators appearing as the
coefficients in the expansions of these fields we have the following
transformation:
\begin{align*}
X_n &\mapsto \frac{2}{n} R_n, \qquad \ol{X}_n \mapsto \frac{2}{n}
\ol{R}_n, \qquad n \neq 0, \\ p_n &\mapsto - U_n, \qquad \ol{p}_n
\mapsto - \ol{U}_n, \qquad n \neq 0, \\ \frac{1}{2}(X_0 + \ol{X}_0)
&\mapsto R_0, \qquad \frac{1}{2}(X_0 - \ol{X}_0) \mapsto \wh{R}_0, \\
(p_0 + \ol{p}_0) &\mapsto p_U, \\ i(p_0-\ol{p}_0) &\mapsto \wt{\om},
\qquad \om \mapsto p_R.
\end{align*}
Thus we see that this transformation exchanges the momentum and the
winding, as expected in $T$--duality. The isomorphism between the two
Hilbert spaces is given by sending $\F_{r+\al,m} \otimes
\ol{\F}_{-r+\al,m}$ to $\wt\F_{r,\al,m}$.

The fermionic Hilbert spaces are the same in the two theories. Hence
we obtain an isomorphism of the full Hilbert spaces of the two
$T$--dual theories.

\ssec{Chiral algebra of the I--model}

In \secref{dual} we defined the deformed model with the action
\eqref{action deformed}, which should be equivalent to the A--model of
$\pone$. The corresponding $T$--dual model is the I--model, which is a
deformation of the theory discussed in the previous section. The
action of this theory is given by formula \eqref{i model1}, and for a
more general toric variety $\PP_S$ it is given by formula \eqref{i
model2}. In this section we will determine the chiral algebra of
integrals of motion of this theory in the sense of
\cite{Zam,FF:laws}. In this context, the I--model is analogous of the
conformal $A_n$ Toda field theory, in which the chiral algebra is the
$\W_n$--algebra (see \cite{FL,FF:laws}). We will show that the chiral
algebra in the I--model corresponding to a toric variety $\PP_S$ is
isomorphic to the space of global sections of the chiral de Rham
complex of $\PP_S$. In order to do this, we will identify the complex
whose zeroth cohomology is this chiral algebra with the complex
introduced by Borisov in \cite{Bor}. The cohomology of this complex is
isomorphic to the cohomology of the chiral de Rham complex of $\PP_S$,
as shown in \cite{Bor}. Thus, the I--model provides a natural link
between Borisov's complex, and hence the chiral de Rham complex of
$\PP_S$, and the sigma model of $\PP_S$.

Consider first the case of $\pone$. The action of the I--model given
by formula \eqref{i model1} is obtained by deforming the action
\eqref{total action} of the free conformal field theory using the
operators $q^{1/2} e^{i U} \pi \ol\pi$ and $q^{1/2} e^{-i U} \pi
\ol\pi$. According to formula \eqref{zam}, for any chiral field $A(z)$
of the free field theory, we have in the I--model
$$
(\pa_{\ol{z}} A)(z,\ol{z}) = q^{1/2} \int e^{i U(w,\ol{z})} \pi(w)
\ol\pi(\ol{z}) dw \cdot A(z) + q^{1/2} \int e^{-i U(w,\ol{z})} \pi(w)
\ol\pi(\ol{z}) dw \cdot A(z).
$$
Therefore the chiral algebra of the I--model is equal to the
intersection of the kernels of the operators $\int e^{\pm i
U(w,\ol{z})} \pi(w) \ol\pi(\ol{z}) dw$ on the chiral algebra $V$ of
the free theory.

The chiral algebra of the free conformal field theory is given by the
direct sum
\begin{equation}    \label{chiral V}
V = \bigoplus_{r \in \Z} \wt\F^{\on{ch}}_{r,r,0} \otimes
\F_{\on{ferm}}.
\end{equation}
Here $\wt\F^{\on{ch}}_{r,r,0}$ is the chiral sector of the Fock
representation $\wt\F_{r,r,0}$ introduced in \secref{T-dual,
operator}. The corresponding chiral fields are normally ordered
products of $\pa_z U(z), \pa_z R(z)$ and their derivatives, as well as
the fields $e^{2 r R_-(z)}$, where $R_-$ is given by formula
\eqref{R-}. The chiral fermionic fields corresponding to vectors in
the chiral fermionic Fock representation $\F_{\on{ferm}}$ introduced
in \secref{hamiltonian} are normally ordered products of $\psi(z),
\pi(z)$ and their derivatives.

We need to find the intersection of the kernels of the operators $\int
e^{\pm i U(w,\ol{z})} \pi(w) \ol\pi(\ol{z}) dw$ on $V$.

Let us write $U(w,\ol{w}) = U_-(w) + U_+(\ol{w})$, where
\begin{align*}
U_-(w) &= \frac{1}{2}(U_0 + p^-_U \log w) + \sum_{n \neq 0}
\frac{U_n}{n} w^{-n}, \\
U_+(\ol{w}) &= \frac{1}{2}(U_0 + p^+_U \log \ol{w}) + \sum_{n \neq 0}
\frac{\ol{U}_n}{n} \ol{w}^{-n},
\end{align*}
and $p^\pm_U = p_U \pm i \wt\omega$. Then it is clear that the kernel
of the operator
$$
\int e^{\pm i U(w,\ol{z})} \pi(w) \ol\pi(\ol{z}) dw = e^{\pm i
  U_+(\ol{z})} \ol\pi(\ol{z}) \int e^{\pm i U_-(w)} \pi(w) dw
$$
on $V$ is equal to the kernel of the operator
$$
S_\pm = \int e^{\pm i U_-(w)} \pi(w) dw.
$$
This allows us to express the chiral algebra of the I--model purely in
terms of modules over a free chiral superalgebra, as we now explain.

Consider the Heisenberg-Clifford superalgebra with the
generators $A_n, B_n, \Phi_n, \Psi_n$, $n$ $\in \Z$, and relations
$$
[B_n,A_m] = n \delta_{n,-m}, \qquad [\Phi_n,\Psi_m]_+ = \delta_{n,-m},
$$
with all other super-commutators being zero. Let $F_{a,b}$ be the Fock
representation of this algebra generated by a vector $|a,b\rangle$
which is annihilated by all generators with $n>0$ and such that
$$
A_0 |a,b\rangle = a|a,b\rangle, \qquad B_0 |a,b\rangle = b|a,b\rangle,
\qquad \Psi_0 |a,b\rangle = 0.
$$
The direct sum $\bigoplus_{a,b \in \Z} F_{a,b}$ is a chiral
algebra. In particular, the field corresponding to the vector
$|a,0\rangle$ is given by the standard formulas
$$
e^{a\int A(z) dz} = \exp \left( a p_A + a A_0 \log z
- a \sum_{n \neq 0} \frac{A_n}{n} z^{-n} \right).
$$
Let us identify the above chiral algebra with our chiral algebra by
the formula
$$
A_n \mapsto - i U_n, \qquad B_n \mapsto 2 R_n, \qquad n \neq 0,
$$
$$
A_0 \mapsto \frac{i}{2} p^-_U, \qquad B_0 \mapsto - R^-_0,
$$
$$
\Phi_n \mapsto \psi_n, \qquad \Psi_n \mapsto i \pi_n, \qquad n \in \Z.
$$

Then our chiral algebra $V$ given by formula \eqref{chiral V} becomes
$F_{0,\bullet} = \bigoplus_{b \in \Z} F_{0,b}$, and the above
operators $S_\pm$ become the operators
\begin{equation}    \label{S+-}
S_\pm = -i \int e^{\pm \int A(z) dz} \Psi(z) dz: \quad F_{0,\bullet} \to
F_{\pm 1,\bullet}.
\end{equation}
Thus, we obtain that the chiral algebra of the I--model is the
intersection of the kernels of the operators $S_+$ and $S_-$ on
$F_{0,\bullet}$.

Now let us compare this with the results of \cite{Bor} (see also
\cite{GMS}). In that paper a complex $C^\bullet$ is constructed such
that $C^0 = F_{0,\bullet}$, and $C^n = F_{n,\bullet} \oplus
F_{-n,\bullet}$, where $F_{n,\bullet} = \bigoplus_{b \in \Z}
F_{n,b}$. The differential is $d = S_+ + S_-$, where $S_\pm:
F_{n,\bullet} \to F_{n \pm 1,\bullet}$ is given by formula
\eqref{S+-}, if $\pm n \geq 0$ and is equal to $0$ otherwise. It is
proved in \cite{Bor} that the $n$th cohomology of this complex is
isomorphic to the $n$th cohomology of the chiral de Rham complex of
$\pone$. The latter vanishes for $n \neq 0,1$, and the $0$th and $1$st
cohomology may be described as modules over the affine Kac-Moody
algebra $\su$ of level $0$ \cite{GMS} (see \remref{su} below).
 
Now we see that this complex, after a change of variables, naturally
appears in the context of the I--model, and hence the A--model of
$\pone$, as anticipated in \cite{Bor}. In particular, we find that the
$0$th cohomology of the chiral de Rham complex of $\pone$ is
isomorphic to the chiral algebra of the I--model, and hence to the
chiral algebra of the A--model associated to $\pone$ (in the infinite
volume limit).

The operators $S_\pm$ are analogues of the screening operators
familiar from the theory of $\W$--algebras (see \cite{FL,FF:laws}). It
is clear from the above formula that they are residues of fermionic
fields. Screening operators of this type have been considered by
B. Feigin \cite{F}.

The generalization of the above computation to the case of the
I--model associated to a toric variety $\PP_S$ is
straightforward. Using a change of variables similar to the one
explained above, we relate the chiral algebra of the I--model
associated to $\PP_S$ to the $0$th cohomology of a complex constructed
in \cite{Bor}. According to \cite{Bor}, the cohomologies of this
complex are isomorphic to the cohomologies of the chiral de Rham
complex of $\PP_S$. In particular, we find that the chiral algebra of
the I--model associated to $\PP_S$ is isomorphic to the $0$th
cohomology of the chiral de Rham complex of $\PP_S$.

\begin{remark}    \label{su}
The toric sigma model carries a chiral $\su$ symmetry with level $0$,
with the generating currents given by the formulas
$$
J^\pm(z) = \Wick (p(z) \pm \psi(z) \pi(z)) e^{\pm X(z)} \Wick \, ,
\qquad J^0(z) = - ip(z).
$$
These formulas can be obtained by a change of variables from the
formulas found in \cite{FF:lmp}, which constitute a special case of
the Wakimoto free field realization. There is also an anti-chiral copy
of $\su$, with the anti-chiral currents given by similar formulas. The
above chiral fields commute with the screening operators and therefore
survive in the deformed theory, and hence we obtain that the A--model
of $\pone$ carries an $\su$ symmetry. It corresponds to the natural
action of the Lie algebra ${\mathfrak s}{\mathfrak l}_2$ on
$\pone$. One can check that these currents are $\ol{Q}(q)$--exact,
where $\ol{Q}(q)$ is the right supercharge discussed in the next
section. \qed
\end{remark}

\ssec{Cohomology of the right moving supercharge}    \label{coh pone}
\label{cohomology}

Now we wish to compute the cohomology of the right moving supercharge
of the I--model and its degeneration. Consider the case of
$\pone$. Recall from \secref{supercharges} that the right moving
supercharge of the I--model is given by the formula
$$
\ol{Q}(q) = \int \left( \ol\psi \pa_{\ol{z}} U d\ol{z} + q^{1/2}
\left( e^{i U} - e^{-i U} \right) \pi dz \right),
$$
and in the $T$--dual variables by
$$
\ol{Q}(q) = \int \left( \ol\psi \ol{p} d\ol{z} + q^{1/2}
\left( e^{i \int P} - e^{- i \int P} \right) \pi dz \right)
$$
(we omit the factor of $-i$ which is inessential for the computation of
cohomology). We wish to compute the cohomology of this operator on the
Hilbert space ${\mc H}$ of our theory that was described in
\secref{hamiltonian}. As the space of functions of the zero mode of
the bosonic fields $X, \ol{X}$ we will take the space of all smooth
functions on $\C^\times$.

We will compute the cohomology of $\ol{Q}(q)$ by utilizing a spectral
sequence corresponding to a $\Z$--bigrading on ${\mc H}$ (we note that
our computation is similar in spirit to the computation in
\cite{W:lg}). The only non-zero degrees are assigned to the fermionic
generators:
$$
\on{deg} \ol\psi_n = - \on{deg} \ol\pi_n = (1,0), \qquad \on{deg}
\pi_n = - \on{deg} \psi_n = (0,1).
$$

Then the first summand of the differential $Q(q)$ has degree $(1,0)$,
while the second summand has degree $(0,1)$. Using the additional
$\Z$--gradings by the eigenvalues of the $L_0$ and $\ol{L}_0$
operators, it is easy to see that the corresponding spectral sequence
converges. The zeroth differential is
$$
\int \ol\psi \ol{p} d\ol{z} = \sum_{n \in \Z} \ol\psi_n \ol{p}_{-n}.
$$
Clearly, it affects only the part of the complex which is generated by
$\ol{p}_n, \ol{X}_n, \ol{\psi}_n, \ol{\pi}_n$. All non-zero modes of
these operators cancel out in the cohomology, and the cohomology
reduces to the cohomology of the operator $\ol\psi_0 \ol{p}_0$ on the
zero mode part of the complex. This operator is the Dolbeault $\ol\pa$
operator, and its cohomology is the space of holomorphic functions on
$\C^\times$ in degree zero, and the other cohomology vanishes. In the
computation that follows we will replace this space by the space of
Laurent polynomial functions on $\C^\times$. This will not affect the
cohomologies.

Thus, we obtain that the first term of the spectral sequence is
given (in the variables of the I--model) by the direct sum
$$
\bigoplus_{m,r \in \Z} \wt\F^{\on{ch}}_{r,r,m} \otimes \F_{\on{ferm}},
$$
where $\wt\F^{\on{ch}}_{r,r,m}$ is the chiral sector of the Fock
representation $\wt\F_{r,r,m}$ introduced in the previous section.
The cohomological gradation corresponds to the fermionic charge
operator. The differential is given by the formula
$$
d =  q^{1/2} \int \left( e^{i U} - e^{-i U} \right) \pi dz.
$$

To relate this complex to the complex considered in \cite{Bor,MS}, we
make the change of variables from the previous section. Then the
complex becomes
$$
C_q = \bigoplus_{r,m \in \Z} F_{m,r}
$$
with the differential $d = q^{1/2} S_+ - q^{1/2} S_-$, where $S_\pm$
are the screening operators from the previous section. Thus, as a
vector space, this complex coincides with the complex $C^\bullet$ from
the previous section, but the differential is different. Indeed, the
differential on $C^\bullet$ was given by formula $S_+ - S_-$ (up to
inessential factors), but by definition $S_+$ acted non-trivially on
$F_{m,r}$ with $m\geq 0$, and by $0$ on $F_{m,r}$ with $m<0$, whereas
$S_-$ acted non-trivially on $F_{m,r}$ with $m\leq 0$, and by $0$ on
$F_{m,r}$ with $m>0$. In contrast, now the differential is defined in
such a way that both $S_+$ and $S_-$ act non-trivially on $F_{m,r}$
with an arbitrary integer $m$. In particular, the cohomological
gradation on $C^\bullet$ introduced in the previous section is now
well-defined only mod $2$.

This new complex is therefore a deformation of the complex
$C^\bullet$, which was previously considered in \cite{Bor,MS}. It was
shown in \cite{MS} that its cohomology is isomorphic to the {\em
quantum cohomology} of $\pone$ (so it is commutative as a chiral
algebra). The cohomology is therefore two-dimensional, and as
representatives of two independent cohomology classes we can take the
identity operator and the operator $q^{1/2}(e^{iU} + e^{-iU})$,
familiar from the Landau-Ginzburg theory.

But what about the complex $C^\bullet$ considered in the previous
section? Following \cite{Bor,MS}, we can interpret it as a a certain
limit of the complex $C^\bullet_q$ when $q \to 0$. To this end, let us
redefine the term $F_{m,r}$ of the complex by multiplying it with
$q^{|m|/2}$. Then the differential $q^{1/2} S_+$, when acting from
$F_{m,r}$ to $F_{m+1,r}, m\geq 0$, will become $S_+$, but when acting
from $F_{m,r}$ to $F_{m+1,r}, m<0$, it will become $q S_+$, and so
will vanish when $q=0$. Likewise, $q^{1/2} S_-$, when acting from
$F_{m,r}$ to $F_{m-1,r}, m\leq 0$, will become $S_-$, but when acting
from $F_{m,r}$ to $F_{m-1,r}, m>0$, it will become $q S_-$, and so
will vanish when $q=0$. Thus, when $q=0$ the complex $C_q$ will
degenerate into the complex $C^\bullet$ considered in the previous
section. Hence its cohomology will become isomorphic to the cohomology
of the chiral de Rham complex of $\pone$ (and so will become much
bigger).

The degenerate complex makes perfect sense as the complex computing
the cohomology of the right moving supercharge in the {\em
perturbative} regime, i.e., without the instanton corrections. Indeed,
in the perturbative regime we consider maps $\Sigma \to \pone$ which
either pass through $0$ or through $\infty$, but not through both
points. We achieve this effect by rescaling the terms of the complex
as described above. According to \cite{W:new,Kapustin}, we should
expect that the cohomology of the right moving supercharge of the
A--model of $\pone$ (which is equivalent to the I--model) is
isomorphic to the cohomology of the chiral de Rham complex of
$\pone$. The above computation confirms this assertion. In addition,
we have also obtained the cohomology of the right moving supercharge
with the instanton corrections and found it to be isomorphic to the
quantum cohomology of $\pone$, using the results of \cite{Bor,MS}.

To summarize, we have a family of complexes $C_q$ depending on a
complex parameter $q$. When $q \neq 0$ the cohomology is
two-dimensional and is isomorphic to the quantum cohomology of
$\pone$, and when $q=0$ the cohomology is isomorphic to the cohomology
of the chiral de Rham complex of $\pone$. Note that we also have a
residual action of the left moving supercharge on this cohomology. For
$q \neq 0$ it simply acts by zero, and so the cohomology of the total
supercharge is the quantum cohomology of $\pone$ as expected in the
A--model of $\pone$. If $q=0$, then it is known (see \cite{MS}) that
the cohomology will be the ordinary (not quantum) cohomology of
$\pone$.

This pattern holds for other Fano toric varieties. Indeed, we can show
in the same way as above that for such a variety $\PP_S$ the
cohomology of the right moving supercharge of the I--model introduced
in \secref{supercharges1} is computed by a complex isomorphic to the
one introduced in \cite{Bor,MS}. The differential obtained from the
supercharges of \secref{supercharges} coincides with the differential
of \cite{Bor,MS}. In fact, we have a family of complexes
parameterized by the K\"ahler cone of $\PP_S$. According to
\cite{MS}, in the case when $\PP_S = \PP^n$ its cohomology is
isomorphic to the quantum cohomology of $\PP^n$. We expect the same to
be true for general Fano toric varieties. This is confirmed by the
computation in \cite{MS} (which uses the results of \cite{Batyrev})
which shows that the cohomology of the total supercharge is isomorphic
to the quantum cohomology of $\PP_S$. Moreover, the cohomology classes
are represented by the elements of the gradient ring of the
superpotential $\wt{W}$ of the I--model. This is what we expect to be
true in the I--model of $\PP_S$, which should be equivalent to the
A--model of $\PP_S$.

In the limit when the parameters of our complex tend to zero, our
complex degenerates. The cohomology of the degenerate complex was
shown in \cite{Bor} to be isomorphic to the cohomology of the chiral
de Rham complex of $\PP_S$. This is again in agreement with the
assertion of \cite{W:new,Kapustin}.

\end{document}